\documentclass[amsmath,amssymb,aps,prl,superscriptaddress,twocolumn,floatfix]{revtex4-2}
\usepackage{graphicx}% Include figure files
\usepackage{dcolumn}% Align table columns on decimal point
\usepackage{bm}
\usepackage{comment}
\usepackage{xcolor}
\usepackage{wasysym}
\usepackage{hyperref} 
\usepackage{physics}
\usepackage{bookmark}
\usepackage{mathtools}
\usepackage{subfigure}
\newcommand{\xCornell}{Department of Physics, Cornell University, Ithaca, NY, USA}
\newcommand{\xEwha}{Department of Physics, Ewha Womans University, Seoul, South Korea}

\definecolor{purple1}{HTML}{a156f0}

\newcommand{\para}{}

\begin{document}

\title{Bionic fractionalization in the trimer model of twisted bilayer graphene}

\author{Kevin Zhang}
\affiliation{\xCornell}
\author{Dan Mao}
\affiliation{\xCornell}
\affiliation{Department of Physics, University of Zurich, Winterthurerstrasse 190, 8057 Zurich, Switzerland}
\author{Eun-Ah Kim}
\email[Corresponding author: ]{eun-ah.kim@cornell.edu}
\affiliation{\xCornell}
\affiliation{\xEwha}
\author{Roderich Moessner}
\affiliation{Max Planck Institute for the Physics of Complex Systems, Dresden, Germany}
%\date{May 2024}

\begin{abstract}
Motivated by the rapid experimental progress in twisted van der Waals materials, we study the triangular trimer model 
as a representative framework for extended Wannier orbitals in twisted bilayer graphene at 
 1/3-filling. This deceptively simple model exhibits a rich suite of complex phases, including unusual excitations exhibiting the physics of fractionalization and fractons. For our investigations, we carry out extensive Monte Carlo simulations using an efficient cluster algorithm. 
 The so-obtained finite-temperature phase diagram reveals a novel polar fluid and an ordered brick-wall phase characterized by fractionally charged $e/3$ excitations with subdimensional lineonic dynamics.
 Notably, we identify a critical trimer liquid phase for the particularly simple model of hard trimers. For this, we derive a new field theory which takes the form of a U(1)$\times$U(1) gauge theory. Its $e/3$ monomers are fractionalized bionic excitations: they carry a {\it pair} of emergent gauge charges, as evidenced by algebraic correlations with two distinct exponents. These field theoretical predictions offer theoretical grounds for 
numerical observations of critical exponents. 
 Our study highlights the triangular trimer model as a new key platform for investigating fractionalization and fractons, 
where trimer liquid bionic monomers can transform into lineons or fractons in proximate phases, and calls for experimental investigations of this physics in twisted van der Waals materials and a broader class of systems with intermediate-range interactions. 
\end{abstract}

\maketitle

\para {\it Introduction.--} The role of constraints is a central theme across a wide variety of physical systems ranging from rigidity theory \cite{Maxwell1864Lond.Edinb.DublinPhilos.Mag.J.Sci.,Kane2014NaturePhys}, hard spheres \cite{Alder1957TheJournalofChemicalPhysics,Boublik1970TheJournalofChemicalPhysics}, water ice \cite{Pauling1935J.Am.Chem.Soc.}, frustrated spin systems \cite{Moessner1998Phys.Rev.Lett.a}, correlated electrons \cite{Gutzwiller1963Phys.Rev.Lett.}, gauge theories \cite{Kogut1979Rev.Mod.Phys.}, and even codes in information theory \cite{Kitaev2003AnnalsofPhysicsa}.
The simplest models capturing the physics of constrained systems, such as (classical and quantum) dimer models \cite{Rokhsar1988Phys.Rev.Lett.a}, have thus been of perennial interdisciplinary interest, encompassing themes such as fractionalization and deconfinement \cite{Moessner2001Phys.Rev.Lett.}, dynamical arrest \cite{Das2005J.Phys.Chem.B,Castelnovo2005Phys.Rev.B}, and entropic ordering \cite{Onsager1949Ann.N.Y.Acad.Sci.}, to name just a few. 
In particular, as a keystone of emergent phenomena, fractionalization \cite{Rajaraman2001arXive-prints} is of significant interest.
Fractionalization involves emergent gauge theories \cite{Castelnovo2012Annu.Rev.Condens.MatterPhys.}, where the fractionalized excitations take the role of electric or magnetic charges of the emergent gauge field.
The possibility of multi-colored (bionic) behavior under the emergent gauge field, as predicted in the context of spin ice and p-orbital bosonic condensate in an optical lattice\cite{Khemani2012Phys.Rev.Bc,Chern2014Phys.Rev.Lett.,Lozano-Gomez2024a}, is particularly intriguing as a phenomenon with no counterpart among fundamental particles.

\para The discovery of systems exhibiting new constrained structures thus has the potential to enlarge our view of the complex static and dynamic phenomena that they engender.
A case in point is the discovery of twisted bilayer graphene (TBG), which has been part of the moir\'e physics revolution in the study of correlated materials currently underway. 
The extended shape of the electronic orbitals in TBG has motivated the study of hard trimer coverings on the triangular lattice.
Such a model has been shown to exhibit extensive ground-state degeneracy at an exactly solvable point \cite{Verberkmoes2001Phys.Rev.Ea,Zhang2022CommunPhysa} and charge fractionalization in an ordered phase away from the solvable point \cite{Mao2023Phys.Rev.Lett.}.
Experimental observations of insulating states at $1/3$ filling  \cite{Cao2018Nature,Xie2021Naturea,Tian2024} suggest that TBG  and other realizations of the trimer model could provide a new avenue to materialize fractionalization based on local constraints. However, little is known about the finite temperature phase diagram or the nature of the fractionally charged monomers. Predictions that can guide experimental pursuits require a comprehensive study of the finite temperature phase diagram and the correlation function of monomers. 

\para In this paper, we take a two-pronged approach of combining large-scale Monte Carlo simulations with effective field theory. The Monte Carlo simulations can accurately capture microscopic considerations, while effective field theory can reveal universal aspects from the simulation results. Employing two global update algorithms, we map out the finite temperature phase diagram, which includes a trimer liquid flanked by a brick-wall and a $\sqrt{3}\times\sqrt{3}$ ordered phase, as well as an adjacent novel polar fluid phase. We investigate the critical properties of the transitions between ordered phases. Furthermore, we study the monomer correlation functions to reveal an algebraic correlation with two exponents defining the trimer liquid phase. Remarkably, an emergent $U(1) \times U(1)$ gauge theory we derive by mapping trimers to three conjoined dimers reproduces the surprising scaling law obtained in Monte Carlo simulations. We comment on how various fractonic phases descend from the trimer liquid with algebraic monomers.

\begin{figure}
    \centering
    \includegraphics[width=.48\textwidth]{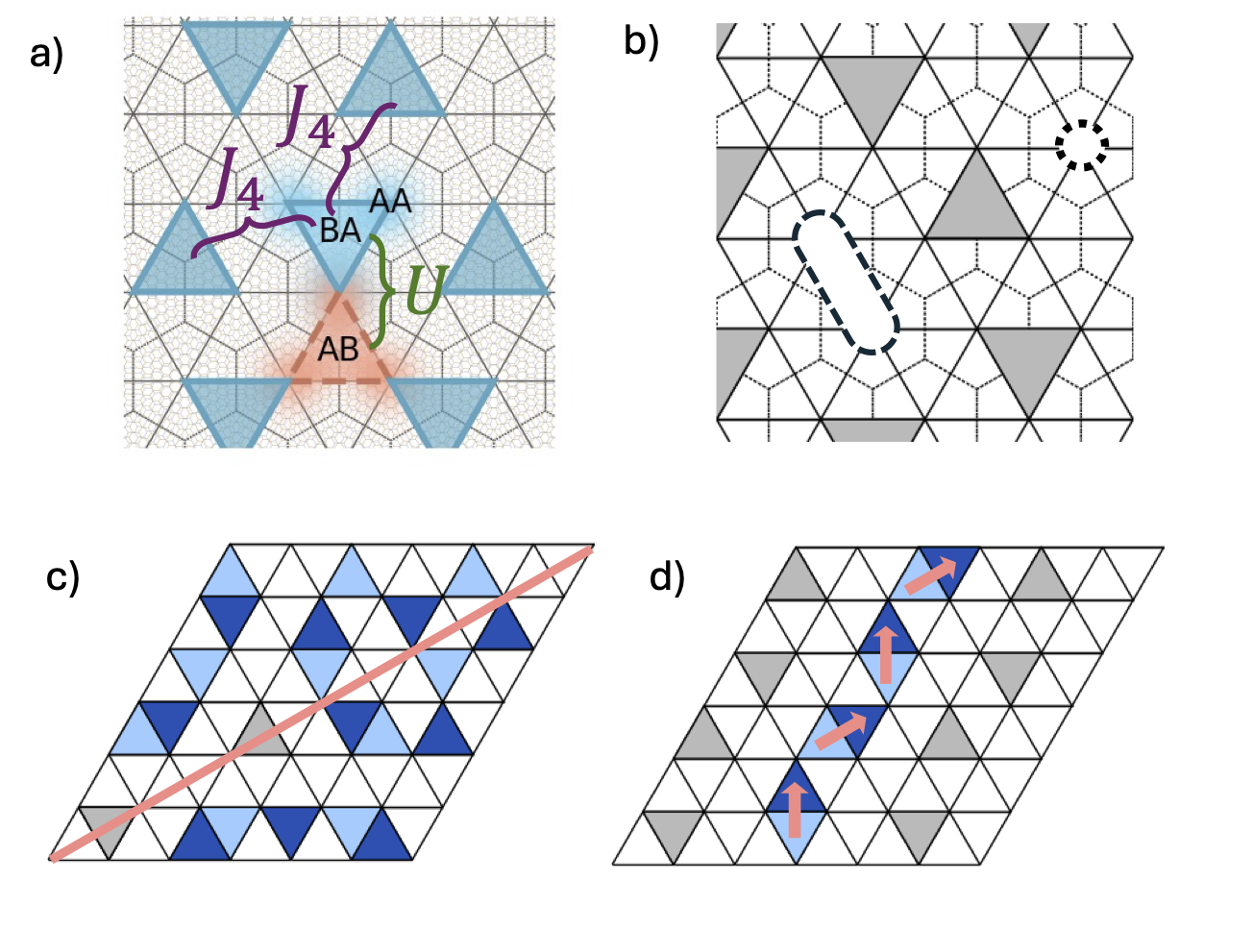}
    \caption{
    \textbf{Overview of the model and Monte Carlo methods.}
    a) 
   TBG Wannier orbitals depicted in fuzzy shades are centered at BA (blue) or AB (red) moir\'e sites, with their lobes extending to three nearby AA sites. We represent the Wannier orbitals by triangles, whose vertices represent three lobes of the Wannier orbitals. 
    The range of $U$ and $J_4$ interactions between trimers is also shown.
    The blue trimers form a trimer covering, a configuration with no overlaps and vacancies. b) A trimer tiling with vacancies. The vacancies are marked with dotted lines. The two vacancies on the left are adjacent and we label them as a dimer with $2/3$ electric charge, and the single vacancy on the right is a monomer with $1/3$ electric charge. c) An example pocket algorithm update, where the light blue and dark blue triangles show the old and new trimer positions that were updated by the algorithm. The pink line marks the symmetry axis. d) An example flip-line update that spans the sample. The light blue and dark blue triangles show the old and new trimer positions that were updated by the algorithm, and the pink arrows mark the flip direction for each trimer.}
    \label{fig:model}
\end{figure}

\para{\it Model.--} 
The extended Wannier orbitals of twisted bilayer graphene \cite{Zhang2022CommunPhysa,Mao2023Phys.Rev.Lett.} can be modeled using trimers centered at the sites of a honeycomb lattice formed by AB and BA moir\'e sites (\autoref{fig:model}(a)).
The trimer vertices at AA moir\'e sites represent the Wannier orbital lobes.
In the strong-coupling limit, the Hamiltonian keeping two leading interaction terms is\cite{Mao2023Phys.Rev.Lett.}
\begin{equation}
\label{eq:model}
    H = \frac{U}{2} \sum_r \left(\sum_{i \in \varhexagon_r} n(i)-1\right)^2 + J_4 \sum_{\langle i,j \rangle_4} n(i) n(j),
\end{equation}
where $n(i)\in \{0,1\}$ is the trimer occupation at AB/BA site $i$.
The cluster charging term of strength $U$  represents Coulomb repulsion between overlapping trimer lobes \cite{Po2018Phys.Rev.X}, while the $J_4$ term represents a further-range interaction as shown in \autoref{fig:model}(a). 
In the rest of this paper, we focus on the filling of one trimer per unit-cell, which amounts to a filling of 1/3. 
We summarize key previous results~\cite{Zhang2022CommunPhysa,Mao2023Phys.Rev.Lett.} for completeness. 
At $J_4=0$, the model reduces to the cluster charging Hamiltonian whose ground states
 are extensively degenerate hard trimer coverings, where every AA site is touched by exactly one trimer with no overlaps (blue trimers in \autoref{fig:model}(a)) \cite{Zhang2022CommunPhysa}.
$J_4$ lifts the degeneracy and drives two distinct ordered states based on its sign: the $\sqrt{3} \times \sqrt{3}$ for $J_4>0$ and the brick-wall phases for $J_4<0$.
Interestingly, Ref.~\cite{Mao2023Phys.Rev.Lett.} found the model in \autoref{eq:model} to support fractionally charged excitations called ``monomers" which carry an electric charge of 1/3 (\autoref{fig:model}(b)). 
Now the key question is whether these monomers can be observable and how they interact with each other, if at all. 
If observable, these monomers will represent a new mechanism for fractionalization in electronic systems that do not rely on Landau levels.

\para {\it Monte Carlo methods.--}
Since the ground state configurations at $J_4=0$ cannot be connected through local moves \cite{Zhang2022CommunPhysa}, an effective global update strategy with short autocorrelation time is essential. To explore the two-dimensional parameter space of $(J_4/U, T/U)$, we combine two global update algorithms: the pocket algorithm~\cite{Krauth2003Phys.Rev.B,Liu2004Phys.Rev.Lett.} and a new global update algorithm we dubbed the  flip-line algorithm. 
The pocket algorithm  \cite{Krauth2003Phys.Rev.B,Liu2004Phys.Rev.Lett.}, which proposes global reconfigurations using  lattice symmetry (see \autoref{fig:model}(c) and SM I A), is rejection-free %\rim{is the acceptance rate really always 100 \% ?}
with respect to both terms of the Hamiltonian (\autoref{eq:model}) and  has a short autocorrelation time.
However, transitions that require a reconfiguration of trimers along a loop are not accessible via the pocket algorithm.
Therefore, we also introduce the global flip-line  algorithm based on the concept of flip-lines introduced in Ref.~\cite{Zhang2022CommunPhysa} (see \autoref{fig:model}(d) and SM I B).
This algorithm is a generalization of the directed-worm algorithm~\cite{Sandvik2006Phys.Rev.Ba} for trimers.
Unlike the pocket algorithm, it is only rejection-free with respect to the $U$ term of the Hamiltonian and can have a long autocorrelation time for large $|J_4|$.
Still, this additional update was necessary for ergodicity, especially at $T=0$.
We combine the pocket and flip-line algorithms by selecting between them randomly with rates of 90\% and 10\%.
Additionally, we use non-reversible parallel tempering \cite{Syed2022JournaloftheRoyalStatisticalSocietySeriesB:StatisticalMethodology} (for a review see SM I C).
Equipped with these tools, we explore the parameter space of $(J_4/U, T/U)$ through Monte Carlo simulations in the canonical ensemble, setting the density of trimers to 1 trimer shared among 3 hexagons. We study monomer correlations in the ground state by removing one or two trimers to create three or six monomers. In all cases, 
we took $5 \times 10^5$ steps to equilibrate the system and performed at least $5\times 10^7$ simulation steps.

\begin{figure*}
\centering
    \includegraphics[width=.96\textwidth]{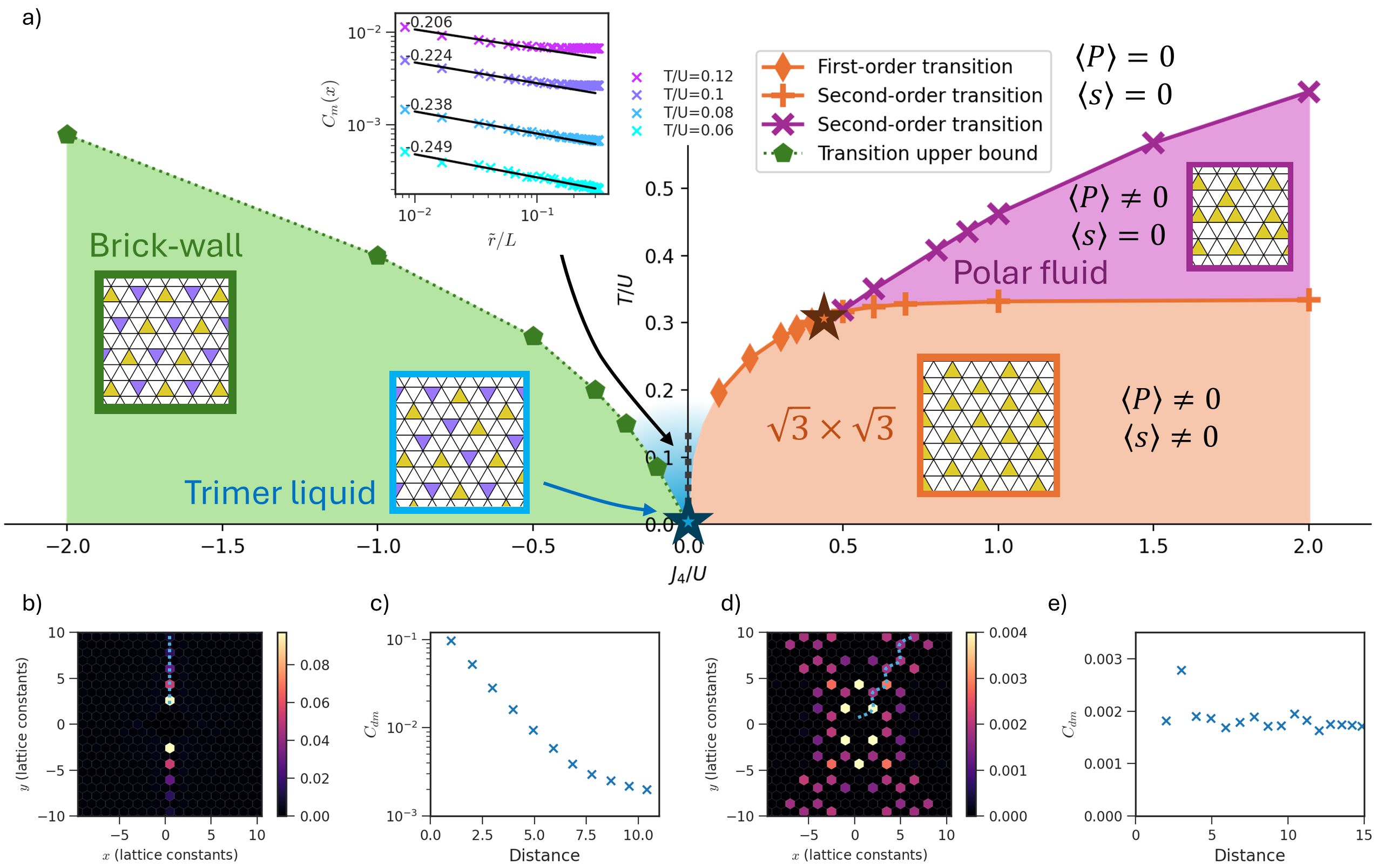}
    \caption{
    \textbf{Phase diagram, algebraic correlation of monomers in the trimer liquid, and restricted mobility of monomers in the ordered phases.}
    a) Phase diagram of the triangular trimer model with interactions.
    Brick-wall and $\sqrt{3}\times\sqrt{3}$ are ordered phases, while the central trimer liquid phase has an extended regime with nearly-critical correlations.
    Characteristic trimer configurations deep in each phase are also shown, with yellow and purple triangles marking up-pointing (AB) and down-pointing (BA) sublattices.
    Cross and plus markers show transition temperatures that were accurately determined from Monte Carlo simulations.
    Pentagon markers show upper-bound on transition temperatures estimated from melting of ordered states (see SM VI).
    Inset: Temperature dependence of short-distance monomer correlations, showing algebraic decay for short distances.
    At distances greater than a temperature-dependent correlation length, the correlations deviate from the power-law line of best fit.
       b) Real-space dimer-monomer correlation function for the ordered $\sqrt{3}\times\sqrt{3}$ phase at $T=0.08$ and $J_4=0.01$. c) Correlation function along the line cut in b) for the $\sqrt{3}\times\sqrt{3}$ phase with an arbitrary distance coordinate. d) Real-space dimer-monomer correlation function for the ordered brick-wall phase at $T=0.15$ and $J_4=-0.05$. Some values have been clipped for clarity. e) Correlation function along the line cut in d) for the brick-wall phase with an arbitrary distance coordinate. 
    }
    \label{fig:pd}
\end{figure*}

\para {\it The Polar Fluid Phase and  $\sqrt{3}\times\sqrt{3}$ Phase --} 
At large and positive $J_4/U$,
 upon cooling down from the  $\mathbb{Z}_6$ symmetric disordered phase, the system first breaks the  $AB$ and $BA$ sublattice symmetry through a second-order transition while preserving the translation symmetry (see SM.~II for detail).  A novel consequence of the extended and anisotropic Wannier orbital is that this sublattice symmetry breaking implies in-plane electric polarization $P \equiv \frac{1}{N}(N_{AB}-N_{BA})$, where $N_{AB}$ and $N_{BA}$ represent the trimer occupation of each sublattice. Hence we name this phase {\it Polar Fluid}. 
 
 \para Upon further cooling, the system undergoes another second-order transition to 
the $\sqrt{3} \times \sqrt{3}$ order phase which breaks both the translational symmetry and the sublattice symmetry. The order parameter for the phase is $s=\sqrt{S(K)/N}$, where $S(K)$ is the structure factor
$S(q) = \frac{1}{N} \left| \sum_r n(r) e^{i r \cdot q} \right|^2$
at $q=K=(4\pi/3, 0)$.
It turns out that the critical exponents remain fixed all along the critical line. Curiously the exponents coincide with those of the three-state Potts model. 
As $J_4/U$ is reduced, the two second-order phase boundaries meet at a tricritical point at 
$J_4/U=0.4$ and $T/U=0.3$. For smaller values of $J_4/U$, we find a direct first order transition between the high-temperature disordered phase and the $\sqrt{3}\times\sqrt{3}$ phase until the system exhibits glassy behavior at smallest $J_4/U$(see SM section II for more detail).
The $\sqrt{3} \times \sqrt{3}$ ordered phase at finite temperature will have thermally excited monomers. 
The calculation of 
dimer-monomer correlations for these thermally excited monomers shows an exponential decay with distance down to a temperature-dependent background density (\autoref{fig:pd}(b-c)), confirming the earlier energetics-based prediction of confinement~\cite{Mao2023Phys.Rev.Lett.}
(\autoref{fig:model}(b)).

\para {\it The Brick-wall phase and lineonic monomers --} We now turn to the $J_4/U<0$ region where the ground state is in the brick-wall phase  \cite{Zhang2022CommunPhysa} that breaks rotational symmetry with order parameter $S(M=(0,2\sqrt{3}\pi/3))$. The brick-wall phase has a residual entropy 
linear in system size and exhibits glassy dynamics. We estimate upper bounds to the transition temperatures, represented by pentagons in \autoref{fig:pd}(a), using the melting temperature of simulations starting from ordered configurations (see SM V).
The monomer-dimer correlation function in the brick-wall phase also exhibits a largely confined character of monomers but in a markedly different manner from that in the $\sqrt{3}\times\sqrt{3}$ phase.
As shown in (\autoref{fig:pd}(d-e)) the correlation function is highly anisotropic, with exponential decay along most directions but showing no sign of decay along two directions. 
This anisotropy signifies the lineonic nature of the monomers in the brick-wall phase that was argued based on ground state energetics in Ref.~\cite{Mao2023Phys.Rev.Lett.}. 
The confirmation of monomer deconfinement along the two easy directions raises the possibility of observation of these lineonic monomers. At the same time, the fact that the movement of monomers cannot relax domain wall boundaries between three orientationally ordered domains makes it challenging to obtain the transition temperature for the brick-wall phase. 

\begin{figure*}
\centering
    \includegraphics[width=.96\textwidth]{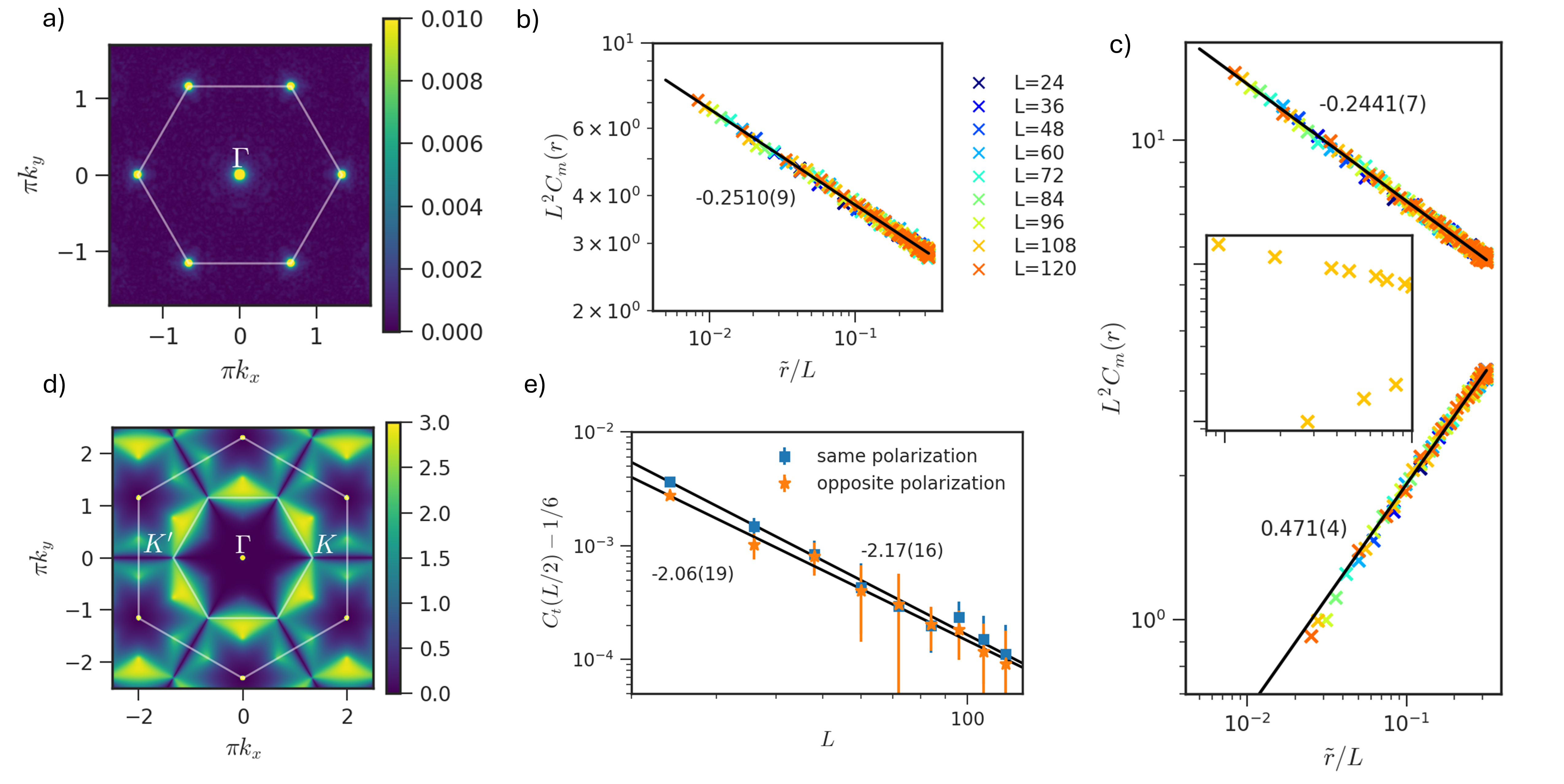}
    \caption{
    \textbf{Numerical results for monomer and trimer correlations in the critical trimer liquid at $T=0$.}
    a) Monomer-monomer correlation function for a system with one trimer vacancy in momentum space, showing isotropic behavior. The white hexagon marks the first Brillouin zone (BZ).
    b) Monomer-monomer correlation function along a cut in real space, for a system with one trimer vacancy. The vertical axis rescaled by $L^2$ to normalize correlation functions. The line of the best fit and the coefficient are also shown.
    c) Monomer-monomer correlation function for a system with two trimer vacancies. The inset shows the oscillating behavior of the correlation function along adjacent sites. The oscillating behavior can be fitted with two distinct power laws.
    d) Trimer structure factor showing two pinch points at $K$ and $K'$; peaks at $\Gamma$ have been clipped for clarity. The inner and outer hexagons denote the first and the second BZ.
    e) Connected trimer-trimer correlation function at inter-trimer distance
     $r=L/2$ between trimers of same and opposite polarization
      as a function of system size.
      A power-law line of best fit is shown. 
    }
    \label{fig:liquid}
\end{figure*}

\para \textit{Critical trimer liquid.--} 
We now turn to the region of the phase diagram that hosts the most surprising discovery: the wedge between two ordered phases anchored on the $J_4/U=0$ point. 
The key question is how the established extensive ground state degeneracy affects the monomer correlation. Although the extensive degeneracy at this point can be obtained exactly through 
Bethe ansatz~\cite{Verberkmoes2001Phys.Rev.Ea}, calculating correlation functions using Bethe ansatz is notoriously difficult.  
We define 
the monomer correlation function  $C_m(\mathbf{r})$
\begin{equation}
\begin{split}
    C_m(\mathbf{r}) &\equiv \langle n_m(\mathbf{r}) n_m(0) \rangle
\end{split}
    \label{eq:monocorr}
\end{equation}
where $n_m(\mathbf{r})$ represents the monomer occupation number at position $\mathbf{r}$. To study monomer correlation at $T=0$, we introduce one or two trimer vacancies which can fractionalize into monomers. 
The structure factor $C_m(\mathbf{q}) \equiv \frac{1}{N} \left| \sum_i e^{-i \mathbf{q} \cdot \mathbf{r}} n_m(\mathbf{r}) \right|^2$ at $T=0$ for a system with one trimer vacancy is isotropic and featureless except for lattice Bragg peaks (see \autoref{fig:liquid}(a)).
However the plot of correlation function $C_{m}(\tilde{r}/L)$ as a function of the conformally mapped distance parameter $\tilde{r}/L \equiv \sin(\pi r/L)/\pi$ (see \autoref{fig:liquid}(b)) reveals the first surprise: a power law decay. This establishes that the monomers form a critical fluid at $J_4/U=T/U=0$. 
The monomer correlations show power-law decay up to a correlation length that decreases with increasing temperature, beyond which they asymptote to a constant and are deconfined.
With a typical energy scale of $U\sim$10~meV in twisted bilayer graphene \cite{Kang2019Phys.Rev.Lett.a}, these critical correlations would persist up to a correlation length of 10 moir\'e lattice constants for $T$ on the order of 10~K, which would be an experimentally accessible regime.

\para The correlation function for a system doped with two trimer vacancies (see \autoref{fig:liquid}(c)) reveal an even bigger surprise. Namely, the $C_{m}(\tilde{r}/L)$ oscillates between two distinct power laws as a function of separation: one decaying and the other growing. To gain insight into this remarkable phenomenon, we probe the trimer correlation function $C_t(\mathbf{r}) \equiv \langle n(0) n(\mathbf{r}) \rangle$, which is equivalent to the electronic correlation function under the mapping to twisted bilayer graphene.
In reciprocal space, the structure factor $C_t(\mathbf{q})$ shows two twofold pinch points at $K$ and $K'$ (\autoref{fig:liquid}(d)).
Each twofold pinch point is indicative of a Gauss' law constraint (see SM IV), and thus the two pinch points reflect the $U(1) \times U(1)$ local symmetry of trimer coverings.
As a function of distance, the connected trimer correlation function $C_t(r)-1/6$ also shows power-law decay
when plotted at a fixed distance $r=L/2$ as a function of system size $L$ (\autoref{fig:liquid}(e)).

\begin{figure*}
\centering
    \includegraphics[width=.96\textwidth]{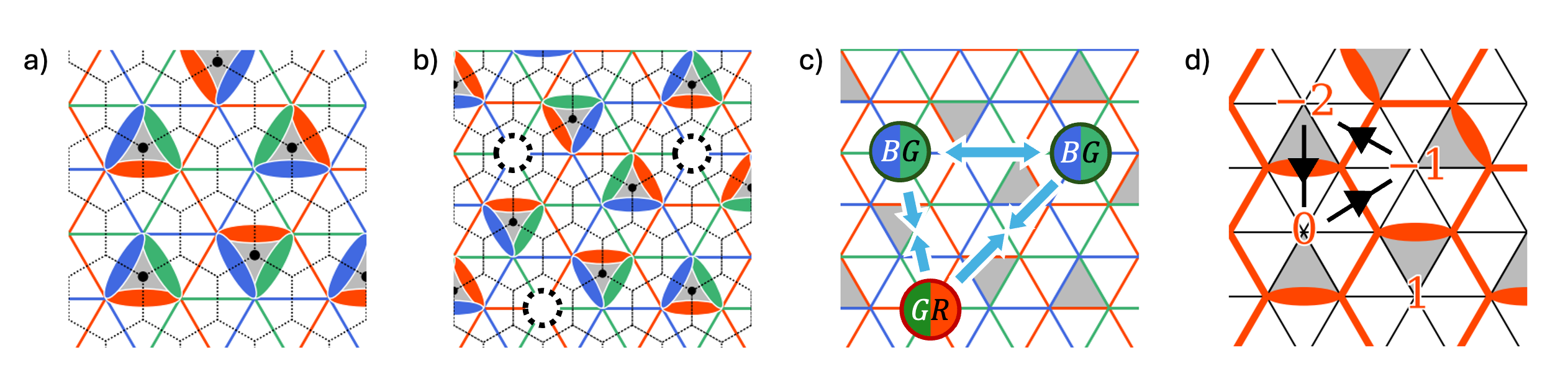}
    \caption{
    \textbf{Conjoined-representation of trimer and derivation of bionic vacancies.}
    a) Tri-coloring of triangular lattice vertices and mapping between the trimers and three dimers.
     b) Monomers as vacancies of the tri-colored lattice.
    c) Bionic nature of monomers (vacancies), which carry charge in two of the dimer colors. Same-color monomers are repulsive, while different-color monomers are attractive.
    d) Height representation of the dimer model of one color (showing red).
    }
    \label{fig:dimer}
\end{figure*}

\para {\it Effective field theory.--}
We now derive an effective field theory that can explain the striking numerical observations of the trimer liquid. 
The aim of the field theory is to reflect hard constraints arising from the cluster-charging term of \autoref{eq:model} and the filling of 1/3 while allowing for a natural description of the monomers as elementary excitations. 
 The key step in the derivation 
 is to map the trimer configuration to a set of three coupled dimer coverings on interpenetrating hexagonal lattices (see \autoref{fig:dimer}(a)): {\it``the conjoined-dimer representation''}. 
 Under this mapping, each vertex of a trimer maps to a site shared by two colored hexagonal lattices.
 The cluster-charging constraint for the trimers translates 
 to the usual constraint for each colored dimer.
In addition, the three dimer coverings are coupled through the constraint that dimers of all three colors combine to form a trimer.
Monomers in the trimer model marked by dashed circles in Fig.~\ref{fig:dimer}(b-c) map to monomers fractionalized from dimers. However, since each monomer site is co-owned by two colored hexagonal lattices the monomers simultaneously fractionalize dimers of two colors. 

\para Now we can leverage the well-established height field representations of a dimer model\cite{Henley1997JStatPhys,Alet2006Phys.Rev.Lett.,Chalker2017TopologicalAspectsofCondensedMatterPhysics} to derive the numerically observed algebraic correlations.
For completeness, we review the height field representation of a dimer model and the monomer correlation function for a bipartite lattice. 
In the absence of monomers, the height fields can be assigned consistently.
An assignment of the height field for a hexagonal lattice starts by choosing one of the sublattices to offer the ``even'' sites for reference.  Then, starting with 
 one dual lattice site to designate $h=0$, 
 follow a counter-clockwise path around a neighboring even site, raising the height field by 2 or lowering it by -1 depending on whether the dual lattice link crosses the original lattice bond with or without a dimer (see Fig.~\ref{fig:dimer}(d)). 
One can then take the continuum limit where the local dimer constraint is automatically satisfied through the curl-free condition $\vec{\nabla}\times\vec{\nabla} h=0$, in the absence of monomers. The resulting field theory is a simple Gaussian field theory with the action~\cite{Chalker2017TopologicalAspectsofCondensedMatterPhysics,Fisher1963Phys.Rev., Youngblood1980PhysRevB}
\begin{equation}
S[h]=\int d{\mathbf r}\frac{K}{2}(\vec{\nabla}h)^2, \quad K=\pi/9.
\end{equation}
Now the dimer field $d({\mathbf r})$ is represented by the $\vec{\nabla}h$'s component along the path, resulting in a power-law dimer-dimer correlation $\langle d({\mathbf r})d(0)\rangle\sim1/r^2$\footnote{Here the sub-dominent contribution from the vertex contribution to the dimer field is ignored. See SM III.}.

\para Monomers violating the dimer constraint introduce vortices, carrying ``magnetic charges'' of $\pm3$ depending on whether they are on ``even'' or ``odd'' sites. A monomer and anti-monomer at neighboring sites form a dimer and get absorbed into the background. Once a dimer is fractionalized into monomers the monomers form a logarithmically interacting ``Coulomb gas".
Following the duality transformation introduced for the XY model 
 a vortex $m({\mathbf r})$ is represented as as pure phase $e^{i3\theta({\mathbf r})}$ associated with the dual bosonic field $\theta({\mathbf r})$ with action $S[\theta]=\int d{\mathbf r}\frac{1}{2K}(\vec{\nabla}\theta)^2$. Using Wick's theorem
the vortex-vortex correlation function can be read off to be
$\langle m({\mathbf r}){m}(0)\rangle\sim r^{{1}/{2}}$ \footnote{A pair of antivortices necessary for net zero vorticity are assumed to be far away.} and the
vortex-antivortex correlation function to be $\langle m({\mathbf r})\bar{m}(0)\rangle\sim r^{-{1}/{2}}$. While the sign of the power-law exponent changes between equal or opposite magnetic charge, the absolute value of the exponent is single valued and determined by the unique stiffness $K=\pi/9$ and the magnitude of the magnetic charge.

\para 
Our conjoined-dimer representation of triangular trimers shown in Fig.~\ref{fig:dimer}(a) has three significant implications. Firstly, the hight fields associated with each colored dimers $h_{R,G,B}$ are coupled 
through the constraint 
\begin{equation}
\vec{\nabla} (h_R+h_G+h_B)=0,
\label{eq:constraint}
\end{equation}
so that three dimers conjoin to form a trimer. 
Secondly, each monomer fractionalized from a trimer is shared by two of the three hexagonal lattice dimer configurations (with vacancies), since every vertex of a trimer is shared by two of the three interpenetrating hexagonal lattices (see Fig.~\ref{fig:dimer}(b)). Hence there are three types of monomors: $m_{GB}$, $m_{RG}$, and $m_{BR}$, depending on the colors of the dimer configuration affected. Thirdly, a physical hole fractionalizes into a three-some of these monomers. Hence it takes three monomers to combine and be absorbed into the trimer background. This is in contrast to a dimer model in bi-partite lattices where two monomers in neiboring lattice are absorbed into the dimer background.   
 
 \para We now use the above implications of the conjoined-dimer representation to derive the power-laws observed in the numerical results. Using the constraint Eq.~\ref{eq:constraint} to integrate out $h_B$ and imposing the symmetry of the underlying lattice requiring permutation symmetry among the three height fields, one can write the effective action up to quadratic order as    
\begin{equation}
    \begin{split}
        S[h_R, h_G] = \int &\frac{1}{2} K \left[(\vec{\nabla}h_R)^2 + (\vec{\nabla}h_G)^2\right]\\
        & + \frac{K}{2} \vec{\nabla}h_R \cdot \vec{\nabla}h_G,
    \end{split}
\end{equation}
where $K=\pi/9$ is the stiffness
 of the dimer model on a honeycomb lattice \cite{Chalker2017TopologicalAspectsofCondensedMatterPhysics}.
Now the effective action can be diagonalized using
orthogonal combinations $h_S\equiv h_R+h_G$ and $h_A\equiv h_R-h_G$, the action is diagonalized:
\begin{equation}
    \begin{split}
        S[h_S, h_A] = \int \frac{1}{2}K_S (\vec{\nabla}h_S)^2 + \frac{1}{2}K_A (\vec{\nabla}h_A)^2,
    \end{split}
\end{equation}
 with $K_S=3K/4=\pi/12$ and $K_A=K/4=\pi/36$ to reveal the two independent $U(1)$ fields manifesting the $U(1) \times U(1)$ symmetry of the trimer liquid. 
In the long-wavelength limit, the trimer-trimer correlations should be the same as the dimer-dimer correlations for any one of the three colors since wherever there is a dimer, there is a corresponding trimer occupation on either side of the dimer. 
Thus, the trimer-trimer correlation should decay as $r^{-2}$ at long distances (see SM III) since the dimer field of any color behaves like a linear combination of $\partial h_S$ and $\partial h_A$. This is consistent with the numerical results in \autoref{fig:liquid}(e).

\para Extending the duality transformation to our conjoined-dimer theory, the three monomer speicies can be represented using the orthogonal dual boson fields as
\begin{equation}
\begin{split}
&m_{GB}(\mathbf{r})\sim  e^{i(3\theta_S(\mathbf{r})-3\theta_A(\mathbf{r}))}\\
&m_{BR}(\mathbf{r})\sim e^{-i(3\theta_S(\mathbf{r})+3\theta_A(\mathbf{r}))}\\
&m_{RG}(\mathbf{r})\sim e^{i6\theta_A(\mathbf{r})}
\end{split}
\label{eq:monomers}
\end{equation}
Since one trimer vacancy fractionalizes into three different monomers, the monomer correlation function in \autoref{fig:liquid}(b) always measures an inter-species correlation. On the other hand, with the introduction of two trimer vacancies, a pair of monomers can belong to the same species or different species~\autoref{fig:liquid}(c). 
The intra-species correlation function is identical to the monomer-monomer correlation in a single hexagonal lattice dimer model, for instance
\begin{equation}
\langle m_{GB}({\mathbf r})m_{GB}({0})\rangle \sim r^{1/2}
\end{equation}
(see SM III for more details). This explains one of the power-laws observed with two trimer vacancies in \autoref{fig:liquid}(c). 

\para By contrast, inter-species correlations necessarily result in a different exponent. Specifically our effective field theory predicts (see SM III)
\begin{equation}
\langle m_{GB}({\mathbf r})m_{BR}({0})\rangle \sim r^{-1/4}.
\end{equation}
This is in remarkable agreement with the second power-law observed in
\autoref{fig:liquid}(b-c).
The different magnitudes of the exponents result from the monomers carrying two $U(1)$ charges, a defining feature of ``bions"\cite{Khemani2012Phys.Rev.Bc}. Such possibility of excitations carrying multiple charges were  first theoretically considered in the context of spin ice on the pyrochlore lattice in Ref.\cite{Khemani2012Phys.Rev.Bc} and in the context of p-band bosons in the diamond optical lattice in Ref.\cite{Chern2014Phys.Rev.Lett.}. The comparison between our effective field theory and the numerical results establishes that $e/3$ charged monomers are bions carrying two additional emergent $U(1)$ charges.

\para {\it Summary and Discussion.--}
We explored the phase diagram and fractionalization of a trimer model on the triangular lattice, motivated by its relevance to TBG at 1/3 fillings. Monte Carlo simulations employing two non-local update algorithms revealed two new phases, in addition to previously identified ordered phases: the polar fluid phase and the trimer liquid phase. In particular, we investigated the critical properties of the transitions between ordered phases and find two second-order phase transitions, one $Z_3$ symmetry-breaking transition with the same universality class as the three-state Potts criticality, and one $Z_2$ symmetry-breaking transition with an unknown universality class. 
Our investigation of various correlation functions established lineonic nature of monomers in the brick-wall phase and algebraic correlation of monomers with two distinct exponents in the trimer liquid phase. By mapping the trimers to three conjoined dimers and extending the height-field formalism, we derived an effective field theory that explains the numerically observed algebraic correlations and the exponents. The effective field theory illucidated bionic origin of the two exponents observed in monomer correlation functions, i.e., the monomers carry 
two distinct emergent $U(1)$ charges in the trimer liquid phase.

\para Despite the apparent simplicity of the cluster charging Hamiltonian with $J_4=0$ in Eq.~\ref{eq:model}, the resulting trimer liquid phase is not only interesting in its own right, but it is at a critical cusp of new rich set of possibilities for fractionalization.  We showed that the monomers change their character from having algebraic correlations in the trimer liquid phase to lineonic mobility in the brick-wall phase with $J_4<0$. The observed glassy dynamics at low temperatures often results from restricted mobility of low-energy excitations (see Ref.~\cite{Ritort2003AdvancesinPhysics} for a review), especially in fractonic phases
\cite{Chamon2005Phys.Rev.Lett., Prem2017Phys.Rev.B, Hering2021Phys.Rev.Ba}.
Moreover, even the confined monomers in $\sqrt{3}\times \sqrt{3}$ ordered phase can turn into isolated fractons or bound fracton pairs ~\cite{Hering2021Phys.Rev.Ba} upon extending the range of interaction to include further neighbors $J_5$ and $J_6$ (see SM section V). 
Alternatively, generalizing the occupation number to a continuous variable and extending the range of interaction up to eighth nearest-neighbor $J_8$ (see SM section V) yield the so-called honeycomb-snowflake model \cite{Benton2021Phys.Rev.Lett.,Yan2023a} with emergent rank-2 $U(1)\times U(1)$ gauge theory with 
gapless fractons. 
The perspective of the cluster-charging model and the trimer liquid phase as the mother of these various descendent fractonic phases offers a fertile ground for future explorations. In particular study of quantum fluctuations can guide quantum simulations of and observation of various fractonic states.

{\it Acknowledgements.--} KZ was supported by NSERC and NSF EAGER OSP\#136036.
DM was supported by the Gordon and Betty Moore Foundation’s EPiQS Initiative, Grant GBMF10436. 
Research at Cornell was supported in part by the National Science Foundation (Platform for the Accelerated Realization, Analysis, and Discovery of Interface Materials (PARADIM)) under Cooperative Agreement No. DMR-2039380. 
This research was supported in part by grant NSF PHY-2309135 to the Kavli Institute for Theoretical Physics (KITP).  Part of this research was performed at the Aspen Center for Physics, which is supported by National Science Foundation grant PHY-160761.
This work was supported in part by
the Deutsche Forschungsgemeinschaft under Grant No. SFB
1143 (Project-ID No. 247310070) and by
the Deutsche Forschungsgemeinschaft under cluster of excellence
ct.qmat (EXC 2147, Project-ID No. 390858490).

\bibliographystyle{apsrev4-2}
\bibliography{refs}

\end{document}

\def \x {\vec{x}}
\def \r {\vec{r}}

\newcommand{\dmadd}[1]{\textcolor{purple}{#1}}
\newcommand{\dmCom}[1]{\textcolor{darkgreen}{\bf {[#1]}}}

\newcommand{\xCornell}{Department of Physics, Cornell University, Ithaca, NY, USA}
\newcommand{\xEwha}{Department of Physics, Ewha Womans University, Seoul, South Korea}

\title{Supplementary Information - Bionic fractionalization in the trimer model of twisted bilayer graphene}

\author{Kevin Zhang}
\affiliation{\xCornell}
\author{Dan Mao}
\affiliation{\xCornell}
\affiliation{Department of Physics, University of Zurich, Winterthurerstrasse 190, 8057 Zurich, Switzerland}
\author{Eun-Ah Kim}
\email[Corresponding author: ]{eun-ah.kim@cornell.edu}
\affiliation{\xCornell}
\affiliation{\xEwha}
\author{Roderich Moessner}
\affiliation{Max Planck Institute for the Physics of Complex Systems, Dresden, Germany}
%\date{May 2024}

\maketitle

\section{I. Monte Carlo methods}
\subsection{A. Pocket algorithm}

\begin{figure}
\centering
    \includegraphics[width=.48\textwidth]{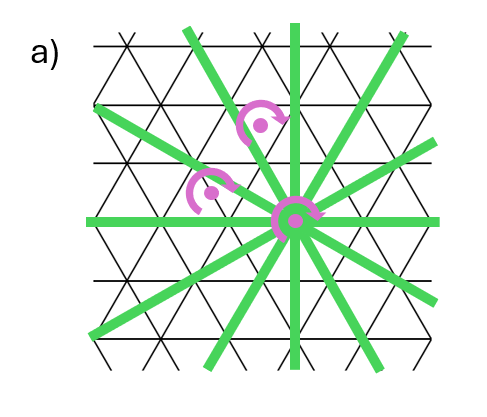}
    \caption{Visualization of translation and rotation lattice symmetries used for pocket algorithm.}
    \label{fig:pocket}
\end{figure}

First, we describe the pocket algorithm \cite{Krauth2003Phys.Rev.B} used in Monte Carlo simulations at zero temperature with $J_4=0$.
We started with reflections about 6 lattice directions centered at triangular lattice vertices and also included $\pi/3$, $2\pi/3$, and $\pi$ rotations around plaquette centers, vertices, and bond centers respectively (\autoref{fig:pocket}).
We implemented the pocket algorithm as follows:
\begin{enumerate}
    \item Choose a symmetry $S$ at random
    \item Initialize sets $P$, $A$
    \item Choose a trimer from the configuration $C$ at random and add it to $P$
    \item While $P$ is not empty:
    \item Remove an element $x$ from $P$, and add $Sx$ to $A$
    \item For all trimers $t$ in $C$:
    \item If $t$ overlaps with $Sx$, add $t$ to $S$
\end{enumerate}
When $S$ is empty, the update is complete.
To generalize this to finite temperatures with interactions \cite{Liu2004Phys.Rev.Lett.}, it is sufficient to change step 7 of the algorithm to: ``Add $t$ to $S$ with probability $\min(1, e^{-\beta \Delta E})$" where $\Delta E = V(t, x) - V(t, Sx)$ and $V(t, x)$ is the two-body interaction between two trimers which includes cluster-charging and $J_4$ terms.
Explicitly, $V(t, x) = nU + mJ_4$, where $n$ is the number of vertices that $t$ shares with $x$, and $m$ is 1 if $x$ and $t$ are fourth-nearest neighbors and 0 otherwise.

\subsection{B. Flip-line algorithm}

\begin{figure}
\centering
    \includegraphics[width=.48\textwidth]{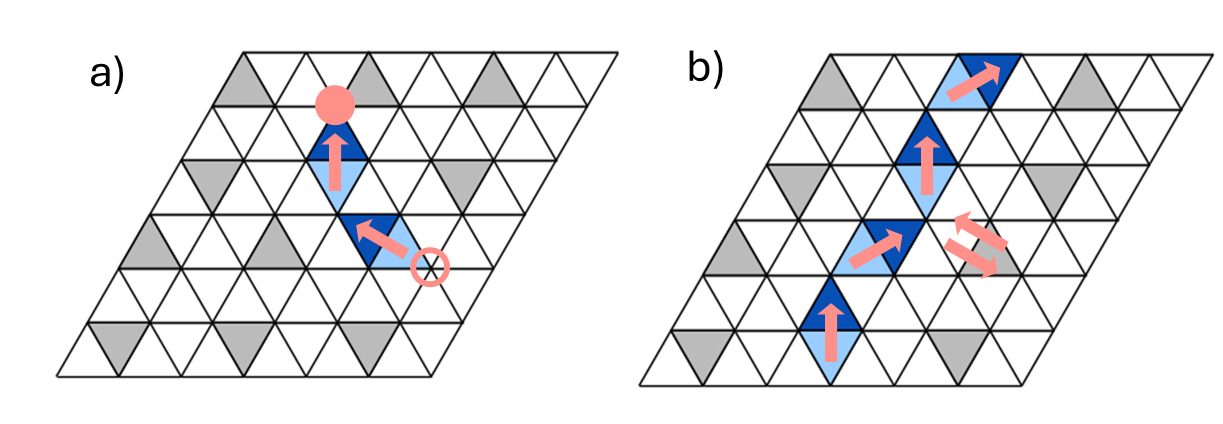}
    \caption{a) A flip-line in progress, two steps in, with the created vacancy and double occupancy marked by empty and filled circles. b) The completed flip-line wraps around the torus and fills in the initial vacancy.}
    \label{fig:flipline}
\end{figure}

The algorithm starts by choosing a random trimer and then flipping it by reflecting it about one of its edges at random.
Next, the newly overlapping trimer is also flipped by reflecting it about the edge opposite the overlap, and so on.
For $T=0$ simulations, the flip-line ends when a vacancy is filled in (\autoref{fig:flipline}(b)).
For finite temperature, it is possible to overlap with multiple trimers after a flip, so we choose only one of them at random to continue the flip-line.
Also, the flip-line is now allowed to terminate at any point after a flip, with probability $P_t=\min(1, e^{-\beta U (n-1)})$, where $n$ is the number of charges on the overlapping vertex after the flip.
This choice of $P_t$ satisfies the detailed balance for our Hamiltonian with $J_4=0$.
For $J_4 \neq 0$, we applied the usual Metropolis-Hastings algorithm using the energy difference of the $J_4$ term of the Hamiltonian.

We now demonstrate that the flip-line algorithm is guaranteed to terminate for trimer coverings.
First, we observe that for any given intermediate double occupancy, the next flipped trimer and flipping direction are uniquely determined.
Also, at $1/6$ filling, the flip network is constrained to a single color of the tri-colored honeycomb lattice, so trimers can only be flipped back and forth along one of their edges.
Thus, the flip-line will eventually close in on itself and fill in the original vacancy (see \autoref{fig:flipline}).

\subsection{C. Optimized parallel tempering}

This section describes the method used to optimize the temperatures of chains used in parallel tempering following the procedure in Ref.~\cite{Syed2022JournaloftheRoyalStatisticalSocietySeriesB:StatisticalMethodology}.
Starting with an ensemble of Monte Carlo chains sorted by temperature from lowest to highest and numbered starting from 1, we alternate between ``odd" and ``even" swaps with a fixed number of Monte Carlo updates occurring on each chain in between.
During a swap, the configuration of each odd (even) numbered chain is proposed to be swapped with the next in sequence.
The swap is accepted with the typical probability $P_{accept} = \min(1, e^{-(\beta_{i+1}-\beta_i)(E_i-E_{i+1})})$.
Furthermore, during an initial calibration period (we took this to be $5 \times 10^5$ Monte Carlo steps), the temperatures of the chains, excluding the first and the last, are optimized.
In order to optimize the temperatures, we first estimate the function $\Lambda(\beta)$, which is a monotonically increasing function called the communication barrier at inverse temperature $\beta$.
For each chain at temperature $\beta_i$, the communication barrier at that temperature is given by $\Lambda(\beta_i) = \sum_{j<i} r(j \to j+1)$, where $r(i \to j)$ is the average rejection probability of swaps between chains $i$ and $j$.
We then approximate $\Lambda(\beta)$ everywhere using linear interpolation between the measured points.
Finally, we redistribute the chain temperatures $\beta_i$ such that $\Lambda(\beta_i) = i/\Lambda(\beta_N)$ where $N$ is the total number of chains.

\section{II. Details of phase transitions}
\begin{figure}
\includegraphics[width=.96\textwidth]{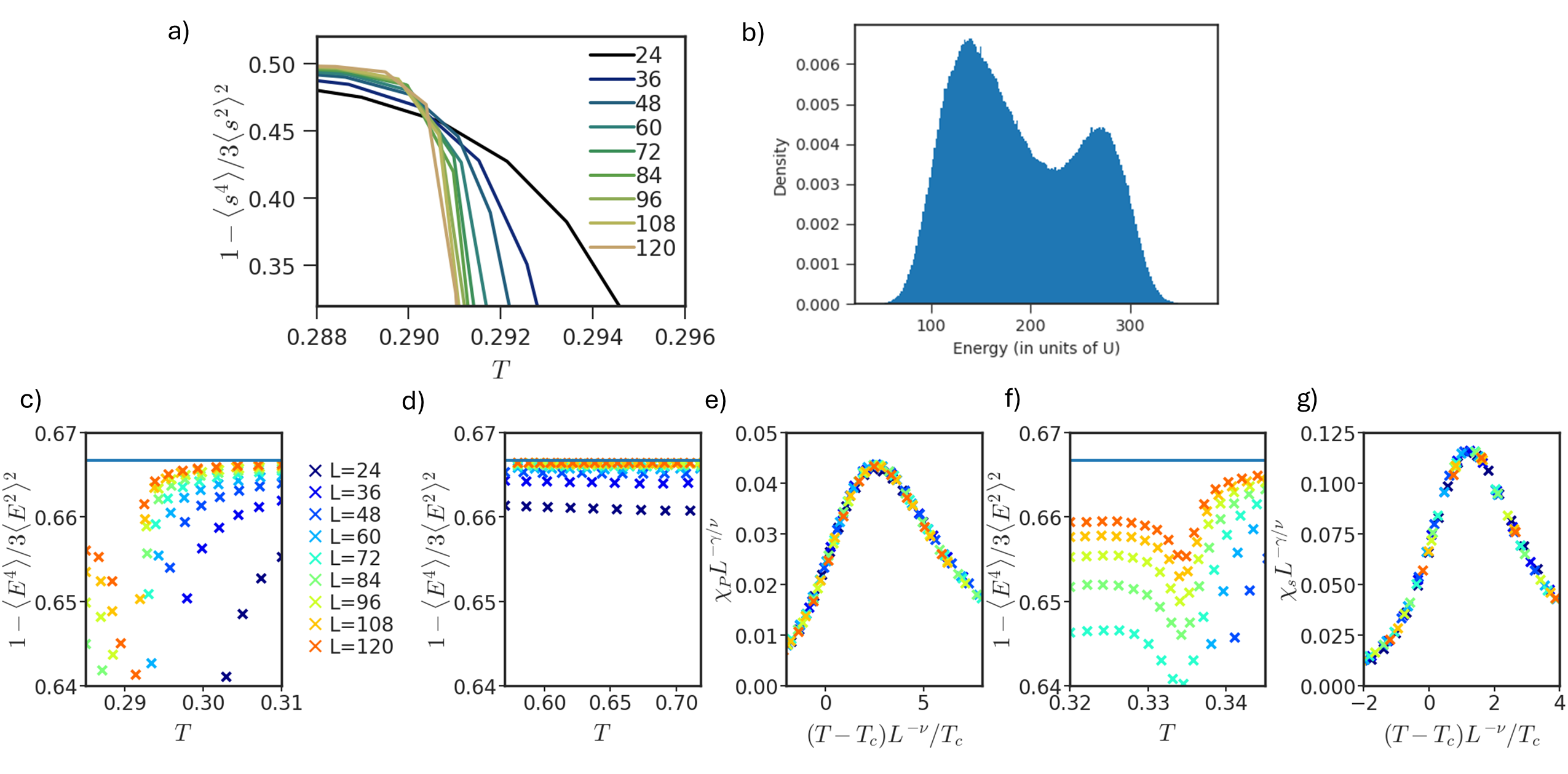}

\caption{
a) Energy histogram near the melting transition between the $\sqrt{3} \times \sqrt{3}$ and disordered phases at $J_4/U=0.35$, showing two peaks.
    b) Binder cumulants for the same transition as a), showing a crossover point which was used to determine the transition temperature.
    c) Fourth-order energy cumulant for the first-order transition out of the $\sqrt{3}\times\sqrt{3}$ phase at $T/U=0.35$ for various system sizes.
    d) Fourth-order energy cumulant for the second-order melting transition of the polar fluid at $T=2$.
    e) Scaling collapse of the susceptibility for the melting transition of the polar fluid at $T=2$.
    f) Fourth-order energy cumulant for the second-order transition between the $\sqrt{3} \times \sqrt{3}$ and polar fluid at $J_4/U=2$.
    g) Scaling collapse of the susceptibility for the transition between the $\sqrt{3} \times \sqrt{3}$ and polar fluid at $J_4/U=2$.
    }
    \label{fig:cumulants}
\end{figure}

In this section we investigate the nature of the phase transitions out of the low-temperature ordered phases for varying $J_4/U$.
Firstly, in order to determine the locations of phase transitions, we uzed Binder cumulants \cite{Binder1981Phys.Rev.Lett.}.
As an example, \autoref{fig:cumulants}(a) shows the Binder cumulant $B_s = 1-\langle s^4 \rangle/2\langle s^2 \rangle^2$ as a function of temperature near the critical temperature for the $\sqrt{3}\times\sqrt{3}$ to disordered transition at $J_4/U=0.35$.
The crossing point of the curves for different lattice sizes is what we determined to be the the transition temperature.

For $J_4/U<0.05$ in the vicinity of the highly degenerate $J_4=0$ point of $J_4/U<0.05$, simulations starting in disordered configurations did not order at any temperature down to $T=0$, instead displaying glassy behavior.
As such, the transition was difficult to probe in numerics.
However, we provide a simple estimate for an upper bound to the melting transition temperature from a free energy calculation: the disordered phase has an entropy lower bound of $\sim 0.257$ per trimer \cite{Verberkmoes2001Phys.Rev.Ea}, while the number of $J_4$ bonds cannot exceed 1.5 bonds per trimer.
The low-temperature transition thus occurs no higher than $T \approx 5.8J_4$.

For the intermediate values of $J_4/U$ we find a direct first order transition between the high-temperature disordered phase and the $\sqrt{3}\times\sqrt{3}$ phase.
The first-order nature of the transitions is supported by two observations.
Firstly, the fourth-order energy cumulant  $1-\langle E^4 \rangle/3\langle E^2 \rangle^2$ 
deviated from the expected value of $2/3$ \cite{Challa1986Phys.Rev.B} for a second-order transition in the thermodynamic limit, even as the system size was increased (\autoref{fig:cumulants}(b)).
Seconly, the energy histogram at $J_4/U=0.35$ (which is near the transition temperature) exhibits two peaks (\autoref{fig:cumulants}(a)) indicating two-phase coexistence \cite{Borgs1992Phys.Rev.Lett.}.

\para %Upon increasing $J_4/U$, the first-order transition line terminates at a tricritical point and splits into two second-order transitions. We estimate the tricritical point to be at roughly $J_4/U=0.4$ and $T/U=0.3$ by considering the intersection of the critical lines.
At larger $J_4/U$, upon cooling down from the  $\mathbb{Z}_6$ symmetric disordered phase, the system first enters the Polar Fluid phase through a second-order transition. %The nature of the transition is probed using energy cumulants.
%breaks the  $AB$ and $BA$ sublattice symmetry while preserving the translation symmetry.  A novel consequence of the extended and anisotropic Wannier orbitals is that this sublattice symmetry breaking implies in-plane electric polarization $P \equiv \frac{1}{N}(N_{AB}-N_{BA})$, where $N_{AB}$ and $N_{BA}$ represent the trimer occupation of each sublattice. Hence we name this phase {\it Polar Fluid}. 
%The ordering is a second-order transition since
The fourth-order energy cumulant of this transition (\autoref{fig:cumulants}(d)) approaches $2/3$. The scaling collapse of the susceptibility $\chi_P \equiv N (\langle P^2 \rangle - \langle P \rangle^2) / T$ at $J_4/U=2$ (\autoref{fig:cumulants}(e)), yielding $T_c=0.625(10)$ and critical exponents $\nu=1.05(13)$ and $\gamma=1.96(10)$, 
which did not appreciably change as $J_4/U$ was tuned.

\para Upon further cooling, the system undergoes another second-order transition to 
the $\sqrt{3} \times \sqrt{3}$ order phase which breaks both the translational symmetry and the sublattice symmetry.
It is also a second-order transition as evidenced by the fourth-order energy cumulant (\autoref{fig:cumulants}(f)) approaching $2/3$.
We performed scaling collapse of the susceptibility $\chi_s \equiv N (\langle s^2 \rangle - \langle s \rangle^2) / T $ at the point $J_4/U$ in \autoref{fig:cumulants}(g), finding $T_c=0.334(1)$ and critical exponents to be $\nu=0.84(5)$, $\gamma=1.46(6)$. It turns out that the critical exponent remain fixed all along the critical line. Curiously the exponents coincide with those of the three-state Potts model. 

\section{III. Height field and effective field theory}
In this section, we derive the effective field theory for the trimer model in detail. First, we review the effective field theory for dimer models on hexagonal lattice via the height field representation. We will briefly review the mapping from dimer model to solid-on-solid (SOS) model, the coarse-grained Gaussian field theory description, and the connection to the Coulomb-gas formalism. From the effective field theory, the correlation function of the dimers and of the monomers can be derived.
Then, we turn to the trimer model and derive the field theory from the perspective of coupled dimer models.

\subsection{Height representation and effective field theory of the dimer model}
\begin{figure}
    \centering
    \includegraphics[width=0.3\linewidth]{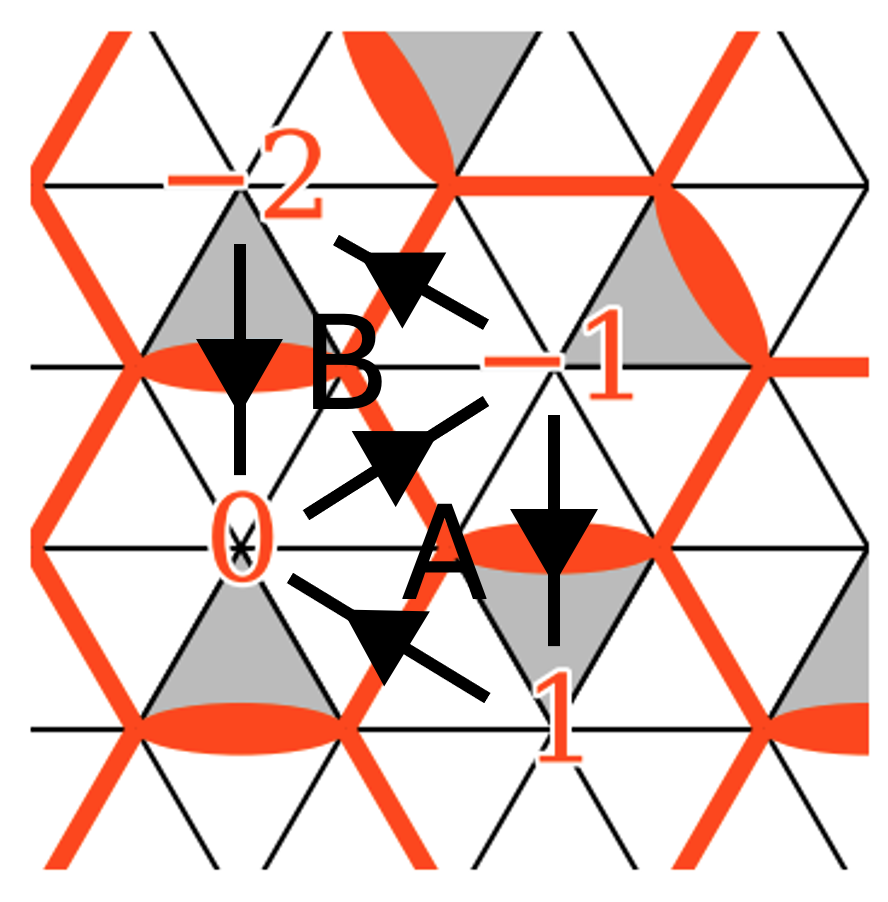}
    \caption{Height mapping for dimer model on hexagonal lattice. The height fields are defined on the plquettes, and the $A, B$ label the two sublattices.}
    \label{fig:sm_dimer}
\end{figure}

The relationship between dimer model on hexagonal lattice, SOS model and antiferromagnetic(AFM) Ising model on triangular lattice was discussed in Ref.\cite{Blote1982J.Phys.A:Math.Gen., Nienhuis1984J.Phys.A:Math.Gen.}. The height variables are integer fields defined on the plaquette centers of the hexagonal lattice (see Fig.\ref{fig:sm_dimer}). Going around the A sublattice clockwise, the height field increases (decreases) by $2$($1$) across an occupied(empty) bond. For the B sublattice, the assignment is the same, but for counterclockwise direction Fig.\ref{fig:sm_dimer}. The height variables are first introduced to describe the height of a crystal-vacuum interface, and the resulting model is called the SOS model. The mapping from the AFM Ising model to the dimer model is a 2-to-1 mapping where the link of the spins with the same sign on the dual triangular lattice maps to the occupied bonds of the dimer model. 

Since for each vertex of the hexagonal lattice, there is always one dimer touching it, the height field is well-defined following the above assignment rules. The height representation would not be very useful if it is merely a rewriting of the dimer configuration. It is argued Ref.\cite{Henley1997JStatPhys} via matching the correlation function from the field theory and the Pfaffian method and later proved Ref.\cite{Kenyon2001Ann.Probab.} that the entropic fluctuation around the most likely configurations can be described by a Gaussian field theory of the coarse-grained height field, where we can simply view the height field as continuous variables. The corresponding effective action can be written as,
\begin{equation}
    S[\phi] = \int d^2 x  \frac12 K (\vec{\nabla} \phi)^2,
    \label{eq:height_action}
\end{equation}
where $\phi$ denotes the coarse-grained height field. The partition function $Z$ that counts all the dimer configurations can, therefore, be viewed as a path integral, namely $Z \sim \int \mathcal{D}[\phi] e^{-S[\phi]}$.

\begin{figure}
    \centering
    \includegraphics[width=0.7\linewidth]{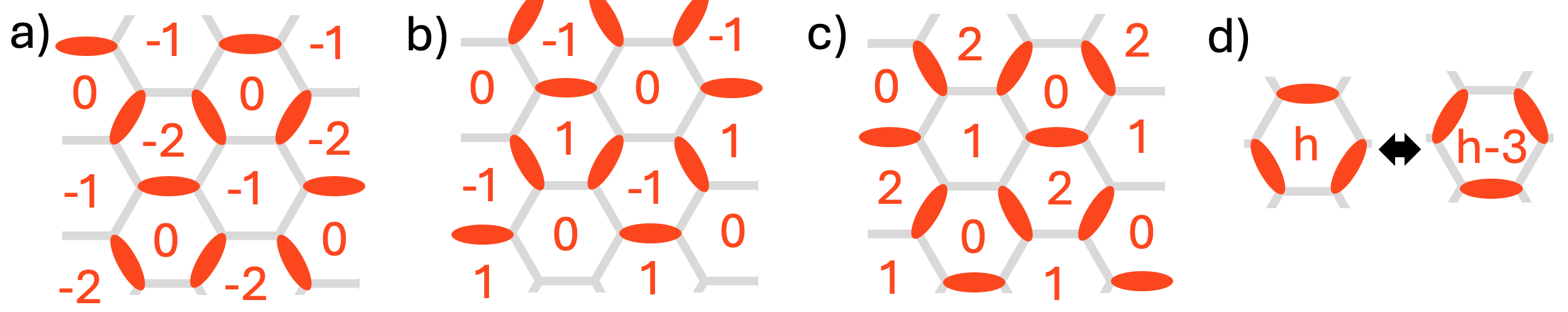}
    \caption{Ideal configurations on the hexagonal lattice and local moves. The three ideal configurations are a),b) and c), related by lattice translation. The local move is illustrated in d). In terms of the local change in height fields, the local move shifts the height of the plaquette by 3.}
    \label{fig:sm_ideal}
\end{figure}

The most likely dimer confiugrations are those with ``flat" height on average and are called ``ideal". On hexagonal lattice, the ideal configurations are those of Kekule pattern, see Fig.\ref{fig:sm_ideal}. The reason for these configurations to be most likely is because they contain a maximal number of local moves of the hexagonal plaquettes. If we start out with one ideal configuration and smoothly deform to another one by performing the local moves, when we go back to the same configuration that we start with, all the height variables increase by $3$. For example, see the height fields in Fig.\ref{fig:sm_ideal} a),b) and c) and apply the local moves to c) to get back to a). This period in height variables defines the so-called compactification radius \cite{Nienhuis1984J.Phys.A:Math.Gen.}. The physical meaning is that any local physical observable defined in terms of the height variable should be invariant under $\phi(\vec{r}) \rightarrow \phi(\vec{r})+3$. Therefore, in the continuum, the physical operators related to the dimer configurations can always be decomposed into superpositions of derivatives of $\phi(\vec{r})$, and $e^{i \frac{2\pi}{3} n \phi(\vec{r})}$, where $n$ is an integer. Now let us consider the dimer occupation field $d(\vec{r})$ in the continuum. By definition, $d(\vec{r})$ is related to the gradient of the height field. On the other hand, since height increased by $3$ gives the same dimer configuration as noted previously, $d(\vec{r})$ should also have overlap with $e^{i \frac{2\pi}{3} n \phi(\vec{r})}$. For the purpose of calculating the exponent of the correlation functions, we can ignore all the numerical coefficients and write
\begin{equation}
   \delta d(\vec{r}) =  d(\vec{r}) - \langle d(\vec{r}) \rangle \sim \partial \phi(\vec{r}) +  e^{i \frac{2\pi}{3} \phi(\vec{r})} + e^{-i \frac{2\pi}{3} \phi(\vec{r})} + ...,
    \label{eq:dimer_op}
\end{equation}
where we ignore all the higher harmonics since their correlation decays faster than the lower one.

A more precise way of finding the relationship between $d(\vec{r})$ and $\phi(\vec{r})$ is via performing coarse-graining on the lattice, where we can define the average height on the vertices of the hexagonal lattice as the average value of the heights on the surrounding plaquettes, and the same relationship as Eq.\ref{eq:dimer_op} can be found (see for example, Ref.\cite{Henley1997JStatPhys, Fradkin2004Phys.Rev.B}). 

Matching the correlation function of the dimers from the Pfaffian solution and the field theory \cite{Kasteleyn1961Physica, Fisher1963Phys.Rev.}, we have $\langle d(\vec{r}) d(0) \rangle \sim 1/r^2$, which gives $K = \frac{\pi}{9}$ on hexagonal lattice.  

The vertex operators $ n(\vec{r}) \equiv e^{i \frac{2\pi}{3} n \phi(\vec{r})}$ is also known as ``electric" charge in the Coulomb gas formalism, which we will touch upon later. The basic idea is that the correlation of the $\phi(\vec{r})$'s decay logarithmically, which is in the form of Coulomb interaction in 2-dimension. 

Now let us turn to the monomers. In terms of the height variables, a monomer can be viewed as a vortex where $\oint \vec{\nabla} h = 3$ going around the monomer. Since the SOS model can be mapped to XY model and ultimately to the Coulomb gas in 2D, the monomer fields can be dealt with using the same methodology as the vortex in the XY model \cite{Kosterlitz1973J.Phys.C:SolidStatePhys.,  Kadanoff1978J.Phys.A:Math.Gen., Jose1977Phys.Rev.B, Nienhuis1984JStatPhys, Domb1987}.

The way to treat the vortex configurations is by introducing the vector fields $\vec{k}$ live on the bonds of the triangular lattice. We let $\vec{k} = \vec{\nabla} h$. Then for a configuration with a number of vortices of strength $m_i$'s living on sites $\vec{R}_i$'s, we have,
\begin{equation}
    \vec{\nabla} \times \vec{k} (\r) = \hat{z} \sum_i m_i \delta(\vec{r}-\vec{R}_i).
\end{equation}
The equation can be solved by noting the similarity to the 2D Gauss Law, and
\begin{equation}
    k_\mu(\r) = -\epsilon_{\mu\nu} \partial_\nu \sum_i \frac{m_i}{2\pi} \text{log}(\frac{|\r - \vec{R}_i|}{a}),
    \label{eq:vortex}
\end{equation}
where $a$ is a UV cutoff and $\epsilon_{xy} = - \epsilon_{yx} = 1$, $\epsilon_{xx}= \epsilon_{yy} = 0$.

Now we can evaluate the free energy of the monomer configurations by simply plugging in the expression for $\vec{k}$ into the Gaussian action,
\begin{equation}
\begin{split}
 S[k] &= \int d^2 x \frac{K}{2} \vec{k}(\r) \cdot \vec{k}(\r)\\
    &= - \frac{K}{4\pi} \sum_{i\neq j} m_i m_j \text{log}(\frac{|\vec{R}_i - \vec{R}_j|}{a}) + \frac{K}{4\pi} \sum_{i} m_i^2 E_{\text{core}},
\end{split}
\label{eq:monomer_int}
\end{equation}
where the first term gives the interaction between different monomers while the second term denotes the energy of the core of the vortex.

Therefore, the interaction between monomers is logarithmic, which is the same as the Coulomb interaction in 2D. Now let us consider the correlation function between the monomers. The difference between the monomer here and the vortex in XY model is that in XY model, the vortices are thermally populated, giving rise to Kosterlitz-Thouless (KT) transition at finite temperature \cite{Kosterlitz1973J.Phys.C:SolidStatePhys.}, while in the dimer model being discussed here, the configurations that contain monomers should be thought of as separate ensembles. The monomer-monomer correlation function should be viewed as the expectation value of the two monomers separated by distance $\vec{r}$, in an ensemble of all possible configurations containing two monomers. And $e^{-S[k]}$ should be viewed as the Boltzmann weight that is proportional to the probability of a certain monomer configuration given by $\vec{k}$.

The correlation between a monomer at position $\vec{r}$ with strength $3$, denoted as $m(\vec{r})$, and an anti-monomer at origin with strength $-3$, denoted as $\bar{m}(0)$ is therefore,
\begin{equation}
    \langle m(\vec{r}) \bar{m}(0)\rangle \sim e^{2 \frac{K}{4\pi} 3 \times (-3) \times \text{log}(r)} \sim \frac{1}{r^{1/2}},
    \label{eq:monomer-corr}
\end{equation}
where we ignore the parts that are independent of the separation of the monomers.

From the perspective of height variables, the dimer and monomer fields are treated very differently. In order to calculate the properties of monomers in a more straightforward fashion, we can use the duality mapping \cite{Jose1977Phys.Rev.B} to relate the monomer to the vertex operator of a dual height field $\theta$. The idea is that one can perform a Hubbard-Stratonovich decomposition of Eq.\ref{eq:monomer_int} based on,
\begin{equation}
    \int D[\theta] e^{-\int d^2 x \left[\frac{1}{2K} (\vec{\nabla} \theta )^2 + f(\r) \theta(\r)\right]} \sim e^{\int d^2 x_1 d^2 x_2 \frac{K }{4\pi}f(\vec{x}_1) f(\vec{x_2}) \text{log}(|x_1 - x_2|)}.
\end{equation}
The upshot is that, if we let $m(\vec{r}) = e^{i 3 \theta(\vec{r})}$, where $\theta(\vec{r})$ field has a Gaussian action,
\begin{equation}
    S[\theta] = \int d^2 x  \frac{1}{2K} (\vec{\nabla} \theta )^2,
\end{equation}
the correlation function of the monomers using the dual-field representation matches the one calculated using monomer-monomer interaction in Eq.\ref{eq:monomer-corr}.

Now, since the dimer fields and monomer fields can all be viewed as vertex operators of scalar fields with logarithmic correlation, we can also view them as ``electric" and ``magnetic" charges interacting with 2D Coulomb interaction, which is the so-called Coulomb gas formalism \cite{Kadanoff1978J.Phys.A:Math.Gen.}. For the magnetic charge, we have already seen that the action of interacting monomers \ref{eq:monomer_int} can be mapped to a Gaussian field theory. For the electric charge, we only need to introduce a source term $e^{i \frac{2\pi}{3} n \phi(\vec{r})}$ to Eq.\ref{eq:height_action} and reverse the above mapping. Then we will find that $n$ is mapped to the electric charge, interacting with each other via logarithmic interaction with a prefactor of $\pi/9K$ instead of $K/4\pi$.   

\begin{figure}
\centering
    \includegraphics[width=.8\textwidth]{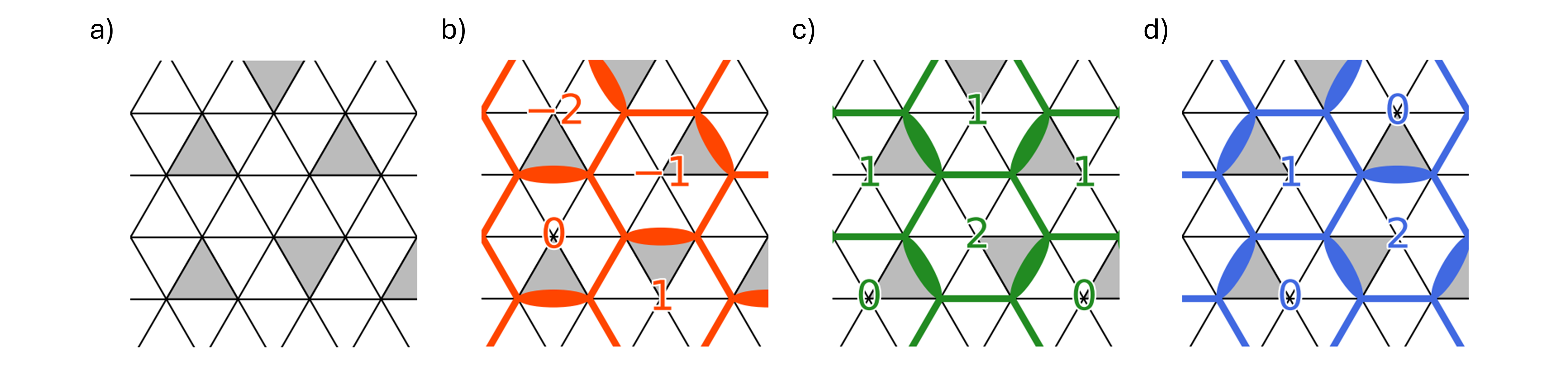}
    \caption{a) A sample trimer configuration. b) Height definitions for the red dimer sublattice. c) Height definitions for the green dimer sublattice. d) Height definitions for the blue dimer sublattice.}
    \label{fig:height}
\end{figure}

\subsection{Effective field theory of the trimer model}
For each dimer model, we can map the dimer configuration to a height field configuration on the dual lattice. Let us denote the height fields as $h_R(x)$, $h_G(x)$ and $h_B(x)$ for the three interpenetrating honeycomb lattice, as shown in Fig.\ref{fig:height}. 
We first study how these fields transform under lattice symmetry and then derive the effective action based on the symmetry perspective.

\subsection{Symmetry transformation of the height fields}
Let us consider the following lattice symmetry. Translation $T_{1,2}$ along the two lattice vectors $\Vec{a}_1 = (1,0)$, $\Vec{a}_2 = (-\frac12, \frac{\sqrt{3}}{2})$, $\mathcal{C}_3$ rotation and inversion $\mathcal{I}$. We choose the center of the rotation and inversion to be the center of the hexagonal plaquette formed by red dimers in Fig.\ref{fig:height}. Under these lattice symmetries, the height fields transform in the following way,
\begin{equation}
    \begin{split}
        T_1&: h_R(\x) \rightarrow h_G(T_1\x), h_G(\x) \rightarrow h_B(T_1\x), h_B(\x) \rightarrow h_R (T_1\x) \\
        T_2&: h_R(\x) \rightarrow h_G(T_2\x), h_G(\x) \rightarrow h_B(T_2\x), h_B(\x) \rightarrow h_R (T_2\x) \\
        \mathcal{C}_3&: h_R(\x) \rightarrow h_R(C_3 \x), h_G(\x) \rightarrow h_G(C_3\x), h_B(\x) \rightarrow h_B (C_3\x)\\
        \mathcal{I}&: h_R(\x) \rightarrow - h_R(-\x), h_G(\x) \rightarrow - h_B(-\x), h_B(\x) \rightarrow - h_G (-\x).
    \end{split}
\end{equation}
The transformation of the height fields under lattice symmetry will constrain the low-energy effective field theory, as we will see in the following.

\subsection{Effective field theory from coupled dimer model}
Since only two out of the three height fields are independent, we can ``integrate out" the third height field and write the effective action in terms of two height fields. From the symmetry transformation properties of the height fields, it is readily seen that the effective action up to quadratic order in terms of the height field is isotropic in $x$ and $y$ direction and can be written as,
\begin{equation}
    \begin{split}
        S[\phi_R, \phi_G] = \int d^2 x \{ &\frac12 K_1 \left[(\vec{\nabla}\phi_R)^2 + (\vec{\nabla}\phi_G)^2\right] + K_2 \vec{\nabla}\phi_R \cdot \vec{\nabla}\phi_G,
    \end{split}
\end{equation}
where $\phi_{R,G}$ are the coarse-grained fields of the height fields $h_{R,G}$, $K_1$ and $K_2$ are parameters to be determined.

If we focus on a single color, say red dimer, the configuration is nothing but the dimer configuration on the honeycomb lattice. Hence, we expect the ``intra-color" behavior of a single color to be the same as the dimer model on a honeycomb lattice, at least to the quadratic level in $\phi$, which gives $K_1 = \pi/9$, as noted in the previous section. The constraint that the three colors of the dimer ``locked" together in order to form a physical trimer will induce ``inter-color" interaction $K_2$ between the dimer models of different colors. Moreover, the locking of the three colors of dimers will give rise to linear constraints of the height fields $h_{R,G,B}$ on the lattice since if the dimer configurations of two colors are fixed, the configuration of the third color is completely determined. In the continuum, due to the permutation symmetry between the three colors, the only linear constraint one can write down is $\phi_R + \phi_G + \phi_B = const$, which gives $\partial(\phi_R + \phi_G + \phi_B) = 0$. 

Now we would like to show that the permutation symmetry and the linear constraint can determine the value of $K_2$. Due to permutation symmetry, the effective action written in $\phi_R$ and $\phi_B$ has the same coefficients $K_1$ and $K_2$ as the one for $\phi_R$ and $\phi_G$, that is,
\begin{equation}
    \begin{split}
        S[\phi_R, \phi_B] = \int d^2 x \left\{\frac12 K_1 \left[(\vec{\nabla}\phi_R)^2 + (\vec{\nabla}\phi_B)^2\right] + K_2 \vec{\nabla}\phi_R \cdot \vec{\nabla}\phi_B \right\}.
    \end{split}
\end{equation}
Plugging in the linear constraint that relates $\phi_B$ to $\phi_G$ and $\phi_R$, we should get back to $S[\phi_R, \phi_G]$. That is, $S[\phi_R,\phi_G] = S[\phi_R, \phi_B = \text{const} -\phi_R - \phi_G]$. Hence, we obtain $K_2 = \frac12 K_1 = \frac{\pi}{18}$.

To obtain the correlation functions, we define the following fields,
\begin{equation}
\left\{
    \begin{split}
        \phi_S &= \phi_R + \phi_G\\
        \phi_A &= \phi_R - \phi_G.
    \end{split}
    \right.
\end{equation}
The action can be diagonalized in the new basis,
\begin{equation}
    S[\phi_S,\phi_R] = \int d^2x \left\{\frac12 K_S (\vec{\nabla}\phi_S)^2 + \frac12 K_A (\vec{\nabla}\phi_A)^2\right\},
    \label{eq:eff_s}
\end{equation}
where $K_S = \frac{1}{2}(K_1 + K_2)$ and $K_A = \frac{1}{2}(K_1-K_2)$.

From the discussion on the dimer model, for vertex operators of the $\phi_{S,A}$ fields, we have,
\begin{equation}
\begin{split}
    \langle e^{i e_S \phi_S(\r)} e^{- i e_S \phi_S(0)} \rangle \sim \frac{1}{r^{\frac{|e_S|^2}{2 \pi K_S} }}, ~
    \langle e^{i e_A \phi_A(\r)} e^{- i e_A \phi_A(0)} \rangle \sim \frac{1}{r^{\frac{|e_A|^2}{2 \pi K_A} }},
\end{split}
\end{equation}
and for the magnetic monopoles of the $\phi_{S,A}$, denoted as vertex operators of the dual fields $\theta_{S,A}$, we have,
\begin{equation}
\begin{split}
    \langle e^{i b_S \theta_S(\r)} e^{- i b_S \theta_S(0)} \rangle \sim \frac{1}{r^{ \frac{K_S|b_S|^2}{2\pi} }},~
    \langle e^{i b_A \theta_A(\r)} e^{- i b_A \theta_A(0)} \rangle \sim \frac{1}{r^{ \frac{K_A |b_A|^2}{2\pi} }}.
\end{split}
\end{equation}

In general, one can consider a generic $n$-point correlator of the vertex operators. From Wick's theorem (for example, see Ref. \cite{Francesco2012} Ch.9), we have,
\begin{equation}
\begin{split}
\langle e^{i e_1 \phi(\r_1)} ... e^{i e_n \phi(\r_n)}\rangle = \delta_{\sum_i e_i, 0}\prod_{i<j} |r_i - r_j|^{\frac{e_i e_j}{2 \pi K}}\\
\langle e^{i b_1 \theta(\r_1)} ... e^{i b_n \theta(\r_n)}\rangle =  \delta_{\sum_i b_i, 0} \prod_{i<j} |r_i - r_j|^{\frac{b_i b_j K }{2 \pi}}
\end{split}
\end{equation}
for a scaler field $\phi(\r)$ with Lagrangian $\frac{1}{2} K (\vec{\nabla} \phi)^2$, whose corresponding dual field is denoted as $\theta(\r)$, with a Lagrangian $\frac{1}{2 K} (\vec{\nabla} \theta)^2$. For a non-vanishing correlator, the charge-neutral condition is required. The correlator is otherwise zero.

The effective field theory allows us to work out the correlation function between monomers and between trimers by mapping them to the orthonormal dimer height fields $\phi_{S,A}$.

First, let us discuss the trimer-trimer correlation. 
Due to the mapping from trimer to dimers, in the long wavelength limit, the trimer-trimer correlation should have the same exponents as the dimer-dimer correlation. Since the trimer field is a bound state of three dimers, $t(\vec{r}) \sim  d_R(\vec{r}) d_G(\vec{r}) d_B(\vec{r})$ and therefore,
\begin{equation}
    \langle t(\vec{r}) t(0) \rangle \sim \langle d_R(\vec{r}) d_G(\vec{r}) d_B(\vec{r}) d_R(0) d_G(0) d_B(0) \rangle = \sum_{ij} \alpha_{ij} \langle \delta d_i(\vec{r}) \delta d_j(0) \rangle + ...,
\end{equation}
where $i,j$ denote $R, G, B$, $\alpha_{ij}$ denotes the product of the expectation value of the dimer occupation for each of the four left-out dimer fields, and the $...$ terms denote higher order terms.

From Eq.\ref{eq:dimer_op}, we have,
\begin{equation}
    \delta d_i(\r) \sim \partial\phi_i(\r) + e^{i \frac{2\pi}{3} \phi_i(\r) } + e^{-i \frac{2\pi}{3} \phi_i(\r) } + ...,
\end{equation}
where we do not specify the exact coefficients since they are not relevant for discussing the exponents and we will follow this convention if not stated otherwise.

It is readily seen that for $i,j \in \{R,G,B\}$,
\begin{equation}
    \langle \partial\phi_i(\r) \partial\phi_j(0) \rangle \sim \langle \partial\phi_S(\r) \partial\phi_S(0) \rangle + \langle \partial\phi_A(\r) \partial\phi_A(0) \rangle
    \sim \frac{1}{r^2},
\end{equation}
and
\begin{equation}
    \langle e^{-i \frac{2\pi}{3} \phi_i(\r) } e^{i \frac{2\pi}{3} \phi_j(0) } \rangle \sim \frac{\delta_{ij}}{r^{8/3}}.
\end{equation}
Therefore, at long distances, we expect $\langle t(\vec{r}) t(0) \rangle \sim \langle \delta d_i(\vec{r}) \delta d_j(0) \rangle \sim 1/r^2$. 

Now, let us consider monomers. For the trimer model, monomers are defined as the plaquette with no occupation. In particular, removing one trimer is equivalent to creating three monomers. 

In terms of the height variable of the coupled dimer models, the monomers are magnetic charges. There are three different types of monomers in the trimer models, denoted as $m_{GB,BR,RG}$ and the corresponding magnetic charges $b_{R, G, B}$ are listed in Tab.\ref{tab:monomer}. Here $m_{GB}$ is the monomer located at the center of the hexagonal plaquette of the red honeycomb lattice in Fig.\ref{fig:height} b), and analogously for $m_{BR}$ and $m_{RG}$. Since $m_{GB}$ violates the dimer constraints for the green and blue dimers but corresponds to opposite sublattices for the two, it carries magnetic charge $\pm 1$ of the two colors, namely $b_G = 3$ and $b_B = -3$, following the convention of the dimer model on hexagonal lattice as discussed previously, similarly for $m_{BR}$ and $m_{RG}$. Note that the bound state of $m_{GB}$, $m_{BR}$ and $m_{RG}$ carries zero magnetic charge, consistent with the fact that the three defects fuse into a physical hole. The corresponding $b_{S,A}$ of magnetic charge of the $\phi_{S,A}$ field (listed in Tab. \ref{tab:monomer}) follows the fact that $b$ denotes the strength of the curl of $\phi$ and therefore the linear transformation between $\phi_{R, G, B}$ and $\phi_{S,A}$ holds for the $b$'s. Namely, $b_S = b_R + b_G$, $b_A = b_R-b_G$.

\begin{table}[h]
    \centering
    \begin{tabular}{|c|c|c|c|c|c|}
    \hline
         &$b_R$ & $b_G$& $b_B$&$b_S$&$b_A$ \\
         \hline
       $m_{GB}$  & 0 &3 &-3&3&-3\\
       $m_{BR}$  & -3 &0 &3&-3&-3\\
       $m_{RG}$  & 3 &-3 &0&0&6\\
       \hline
    \end{tabular}
    \caption{Magnetic charge of the monomers}
    \label{tab:monomer}
\end{table}

Now, let us consider the correlation between two monomers of the trimer model. Since the physical hole always contains three monomers, the correlation functions of the monomers are always calculated with the fixed position of some background monomers and varying the relative position of two monomers of interest.

We first consider the correlation function between two monomers of different types, say $m_{GB}$ and $m_{BR}$, with a background $m_{RG}$ whose position is fixed to be at $\vec{w}$, 
\begin{equation}
\begin{split}
        &\langle m_{GB}(\r) m_{BR}(0) m_{RG}(\vec{w}) \rangle\\
        \sim& \langle e^{i (3 \theta_S(\r) - 3 \theta_A(\r))}e^{i (-3 \theta_S(0) -3\theta_A(0))} e^{6 i\theta_A (\vec{w})}\rangle \\
        \sim& \frac{1}{r^{3/8}} \frac{1}{r^{-1/8}|\r-\vec{w}|^{1/4} w^{1/4}}\\
        \propto& \frac{1}{r^{1/4}},
\end{split}
\end{equation}
where the last line is obtained by choosing a fixed $w$ such that $|w| \gg |r|$ and we only care about the $r$ dependence. Same exponents can be obtained by considering correlators between $m_{GB}$, $m_{RG}$ and $m_{BR}$, $m_{RG}$, due to the permutation symmetry of $A$, $B$ and $C$.

For monomers of the same types, we need to at least dope two holes into the system. Therefore we consider the correlation between two $m_{GB}$'s, with background monomers of two $m_{BR}$'s and two $m_{RG}$'s. Analogous to the previous case, we also choose the coordinate of the background monomers to be far away from the monomers that we are interested in. Therefore, 
\begin{equation}
\begin{split}
        &\langle m_{GB}(\r) m_{GB}(0) m_{BR}(\vec{w}_1) m_{BR}(\vec{w}_2) m_{RG}(\vec{y}_1) m_{RG}(\vec{y}_2) \rangle\\
        \sim& \langle e^{i (3\theta_S(\r) - 3\theta_A(\r))} e^{i (3\theta_S (0) - 3\theta_A(0))} ... \rangle \\
        \propto& r^{1/2},
\end{split}
\end{equation}
where $...$ is the product of the background monomer fields.

These monomer-monomer correlation functions are simulated using the classical MC and the simulated exponents are shown in Fig. 3 d) and e) in the main text.

\section{IV. Pinch points and $U(1)$ gauge theory}

\begin{figure}
    \centering
    \includegraphics[width= 0.3
    \textwidth]{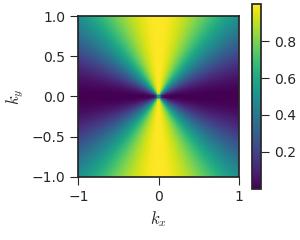}
    \caption{$\langle E_x(\vec{q})E_x(\vec{q})\rangle$ for the low-energy momentum space form of the Gauss law constraint, showing how the $U(1)$ constraint arises in a pinch point.}
    \label{fig:pinch}
\end{figure}

In this section, we review the origin of the two-fold pinch points.
The two-fold pinch point arises when there is a Gauss law constraint at low energy.
To see this, note that in momentum space, the Gauss law gives $\vec{q}\cdot \vec{E}(\vec{q}) = 0$. Therefore, $\vec{E}(\vec{q})$ is purely transverse and the correlation function behaves as
$\langle E_i(\vec{q}) E_j(\vec{q})\rangle \sim  \delta_{ij} - q_i q_j /|q|^2$. This projection to the transverse component gives two-fold pinch points when we go around $q=0$ since when $q_i = 0$,  $\langle E_i(\vec{q}) E_i(\vec{q})\rangle $ vanishes (see \autoref{fig:pinch}).

\section{V. Extensions of cluster-charging model and fractons}
\subsection{Classical Kagome spin model}
By introducing fifth and sixth nearest-nerighbor interactions $J_5$ and $J_6$, and letting $J_4=J_5$ and $J_6>0$, the system has ``subextensive" degeneracy arising from energetically free domain walls in the $\sqrt{3}\times \sqrt{3}$ phase.

%This case can be mapped to the kagome lattice model studied in Ref.~\cite{Hering2021Phys.Rev.Ba}, hosting isolated fractons and bound fracton pairs.

\subsection{Honeycomb snowflake model}
\begin{figure}
    \centering
    \includegraphics[width= 0.5\textwidth]{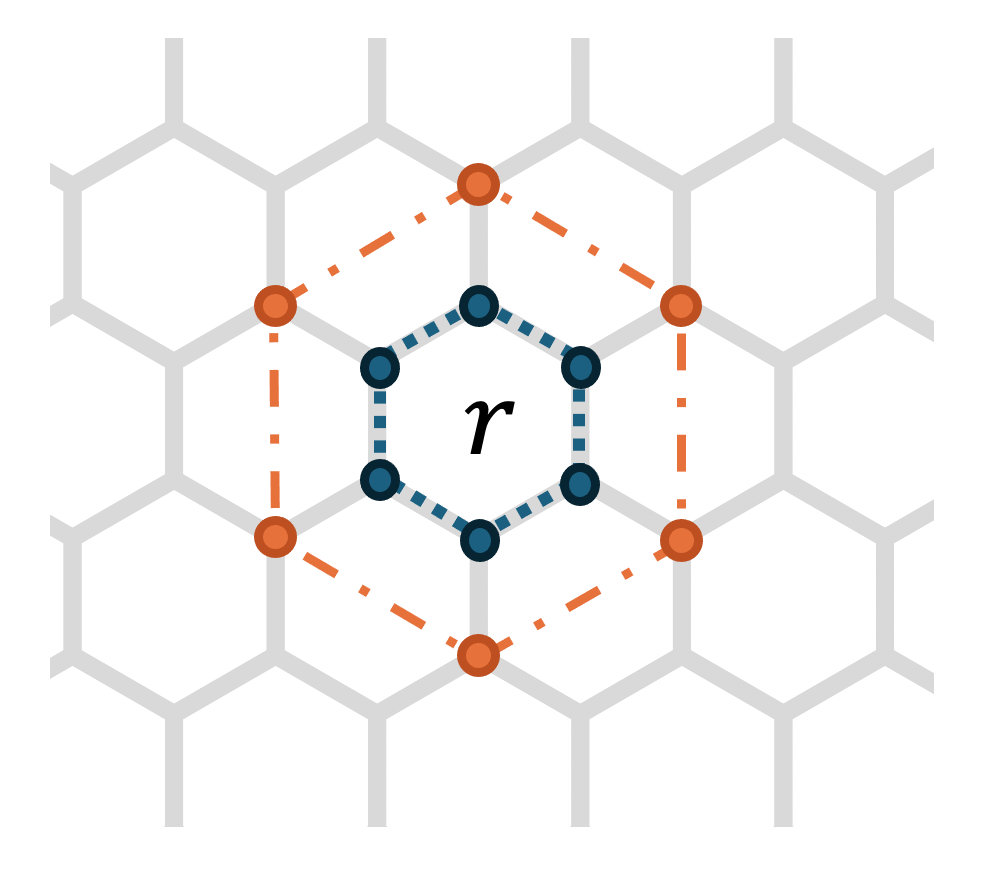}
    \caption{Hamiltonian for the honeycomb snowflake model. The blue dotted plaquette denotes $\varhexagon_r$, and the orange dashed plaquette denotes $\varhexagon'_r$ in Eq.\ref{eq:hs}, where the $n(i)$'s live on the sites of the corresponding plaquettes marked by the blue and orange dots.}
    \label{fig:hs}
\end{figure}

We review the honeycomb snowflake model studied in Ref.\cite{Yan2023a}. The model is defined on the honeycomb lattice, where the constraints can be written as satisfied by the ground state of the Hamiltonian $H_{hs}$, where
\begin{equation}
     H_{hs}=\frac{U}{2} \sum_r \left[\left(\sum_{i \in \varhexagon_r} n(i)\right)+ \gamma \left(\sum_{i' \in \varhexagon'_r} n(i')\right) \right]^2,
     \label{eq:hs}
\end{equation}
and the $n(i)$'s are treated as continuous variables instead of discrete variables, as studied in the main text. The two different honeycomb plaquettes in the summation are the closest and second closest honeycomb plaquettes around the center labeled by $r$ (see Fig.\ref{fig:hs}). Note that when $\gamma = 0$, $H_{hs}$ is the continuous version of the cluster-charging constraints. When $\gamma \neq 0$, we can also write $H_{hs}$ as a summation of further range density-density interaction terms, and the $\gamma$ term introduces up to the eighth nearest neighbor interaction. 

By treating $n(i)$'s as continuous, the structure factor can be readily obtained by Fourier transformation of $H_{hs}$ \cite{Yan2023a}. Let us discuss two special limits: $\gamma = 0$ and $\gamma = 1/2$.
\begin{itemize}
    \item $\gamma = 0$: 
    In Fourier space, we can write $H_{hs}$ as,
    \begin{equation}
        H_{hs}(\gamma = 0) = \frac{U}{2}\sum_k |\hat{f}^T(\vec{k}) \hat{n}(\vec{k})|^2,
    \end{equation}
    where $\hat{n}(\vec{k}) \equiv \left(n_A(\vec{k})~ n_B(\vec{k})\right)^T $ denotes the Fourier transform of $n(i)$ on A, B sublattices and $\hat{f}(\vec{k})=  \left(f_A(\vec{k})~ f_B(\vec{k})\right)^T$ where
\begin{equation}
\begin{split}
    f_A(\vec{k}) &= 1+e^{-i \vec{k}\cdot \vec{a}_1} +e^{-i \vec{k}\cdot (\vec{a}_1+\vec{a}_2)}\\
    f_B(\vec{k}) &= 1+e^{-i \vec{k}\cdot \vec{a}_1} +e^{i \vec{k}\cdot \vec{a}_2}
\end{split}
\end{equation}
and $\vec{a}_1 = (1,0)$, $\vec{a}_2 = (-1/2, \sqrt{3}/2)$ are the lattice vectors.

The emergent Gauss Law can be readily seen from the structure of $\hat{f}(\vec{k})$. In the Brillouin zone, $\hat{f}(\vec{k})$ vanishes when $\vec{k}$ is at $K$ or $K'$ point. Let us expand $\hat{f}(\vec{k})$ around $\vec{K} = (\frac{4\pi}{3\sqrt{3}},0)$,
\begin{equation}
    \hat{f}(\vec{q} + \vec{K}) = \frac{\sqrt{3}}{2} e^{i \frac{\pi}{3}}(q_x + i q_y ~ q_x -i q_y ) + O(q^2). 
\end{equation}
Therefore, for $n(\vec{k})$ to satisfy the constraints, we have,
\begin{equation}
    (q_x + i q_y)n_A(\vec{q}+\vec{K}) + (q_x - i q_y)n_B(\vec{q}+\vec{K}) = 0,
\end{equation}
which give rise to a Gauss Law $\vec{q} \cdot  \vec{E}_{\vec{q}} = 0$ in the long wavelength limit if we define
\begin{equation}
    \vec{E}_{\vec{q}} \equiv \left(n_A(\vec{q}+\vec{K}) + n_B(\vec{q}+\vec{K}), ~ i n_A(\vec{q}+\vec{K}) - i n_B(\vec{q}+\vec{K}) \right),
\end{equation}
similarly for $n(\vec{k})$ near $K'$ point.  Although it seems that the electric field is complex, we note that $n^*(\vec{k}) = n(\vec{k})$ so the two ``complex" Gauss Law at $K$ and $K'$ points can combine into two independent Gauss Laws for two real electric fields.

    \item $\gamma = \frac12$:
Analogous to the $\gamma = 0$ case, in Fourier space, we can write $H_{hs}$ as,
    \begin{equation}
        H_{hs}(\gamma = \frac12) = \frac{U}{2}\sum_k \left|\left(\hat{f}^T(\vec{k}) + \frac12 \hat{g}^T(\vec{k}) \right) \hat{n}(\vec{k})\right|^2,
    \end{equation}
    where $\hat{g}(\vec{k})=  \left(g_A(\vec{k})~ g_B(\vec{k})\right)^T$ and
\begin{equation}
\begin{split}
    g_A(\vec{k}) &= e^{-i \vec{k}\cdot \vec{a}_2}  + e^{-i\vec{k}\cdot(2 \vec{a}_1+ \vec{a_2})}+ e^{i \vec{k}\cdot \vec{a}_2} \\
    g_B(\vec{k}) &= e^{i \vec{k}\cdot (\vec{a}_1 + \vec{a}_2)}  + e^{-i\vec{k}\cdot(\vec{a}_1+ \vec{a_2})} + e^{-i \vec{k}\cdot (\vec{a}_1-\vec{a}_2)}.
\end{split}
\end{equation}
$\hat{g}(\vec{k})$ vanishes at $K$ and $K'$ points as well as $\hat{f}(\vec{k})$ and we can expand $\hat{f}(\vec{k}) + \frac12 \hat{g}(\vec{k})$ around $K$ point,
\begin{equation}
    \hat{f}(\vec{q}+\vec{K}) + \frac12 \hat{g}(\vec{q}+\vec{K}) = \frac38 e^{i \frac{\pi}{3}} (- q_x^2+ 2i q_x q_y+ q_y^2~ - q_x^2- 2i q_x q_y+ q_y^2)+ O(q^3),
\end{equation}
which corresponds to
\begin{equation}
    (- q_x^2+ 2i q_x q_y+ q_y^2)n_A(\vec{q}+\vec{K}) + (- q_x^2- 2i q_x q_y+ q_y^2)n_B(\vec{q}+\vec{K}) = 0,
\end{equation}
for $n(\vec{k})$ to satisfy the constraints. Therefore, if we further define a rank-2 electric field $E_{\alpha \beta}$ such that
\begin{equation}
\left(
\begin{array}{cc}
    E_{xx} & E_{xy} \\
    E_{yx} & E_{yy}
\end{array}
\right)
= \left(
\begin{array}{cc}
    -n_A(\vec{q}+\vec{K}) - n_B(\vec{q}+\vec{K}) & in_A(\vec{q}+\vec{K}) - i n_B(\vec{q}+\vec{K}) \\
    in_A(\vec{q}+\vec{K}) - i n_B(\vec{q}+\vec{K})  & n_A(\vec{q}+\vec{K}) + n_B(\vec{q}+\vec{K})
\end{array}
\right),
\end{equation}
the electric fields satisfy the Gauss Law for a rank-2 gauge theory, $\partial_\alpha \partial_\beta E_{\alpha \beta} =0$.

\end{itemize}

\section{VI. Glassy dynamics and hysteresis}

\begin{figure}
    \centering
    \includegraphics[width= 0.9\textwidth]{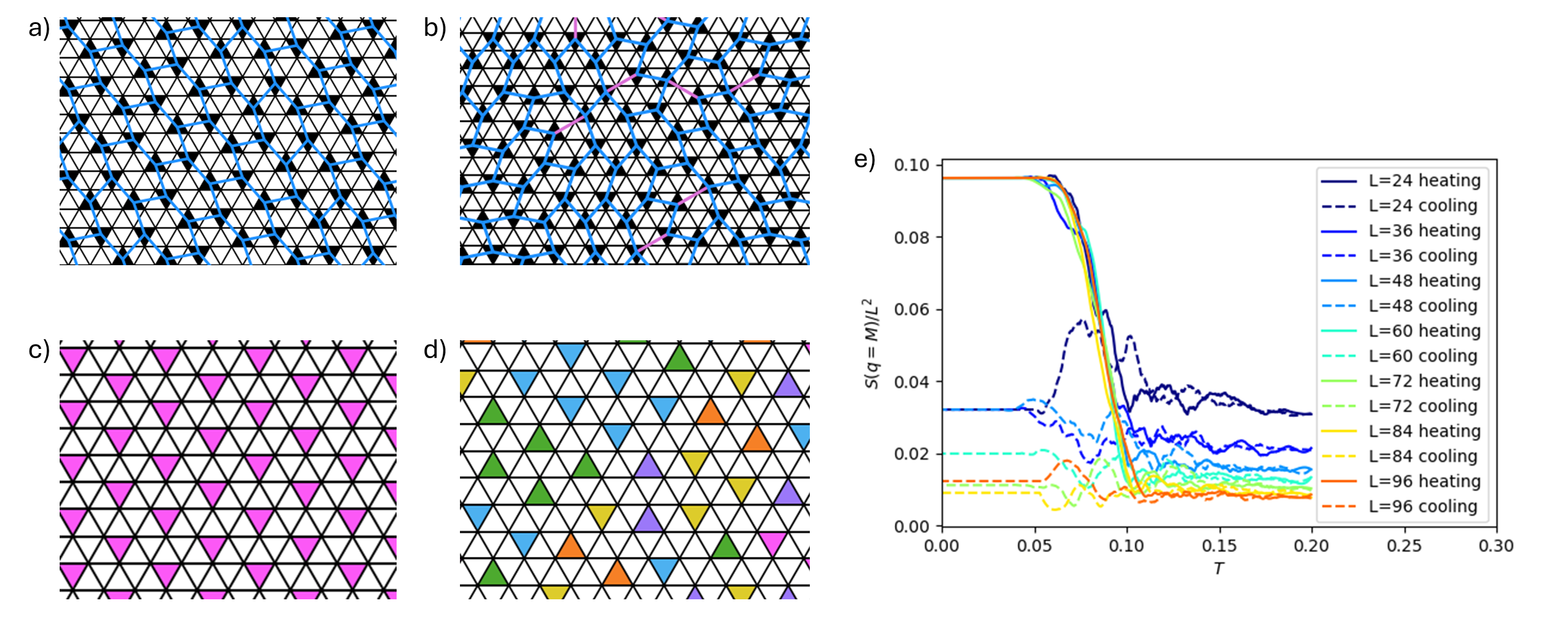}
    \caption{Glassy configurations and hysteresis in heating-cooling loop used to determine transition upper bound. a) Brick-wall monocrystal. Blue lines mark $J_4$ bonds. b) Brick-wall domains obtained from a $48\times 48$ simulation starting from a disordered configuration at $T/U=0$ and $J_4/U=-0.1$. Purple lines mark broken $J_4$ bonds leading to an energy cost. c) $\sqrt{3} \times \sqrt{3}$ monocrystal. d) $\sqrt{3} \times \sqrt{3}$ domains obtained from a $48 \times 48$ simulation starting from a disordered configuration at $T/U=0$ and $J_4/U=0.01$. Trimer colors mark the six different possible $\sqrt{3} \times \sqrt{3}$ registries. e) Heating-cooling hysteresis loop of the structure factor at $q=M$ obtained from a simulation at $J_4=-0.1$. Starting from an ordered configuration, the Monte Carlo simulation was carried out starting from $T=0$ and slowly increasing temperature up to $T=0.2$ (solid lines), at which point the brick-wall order melted. The temperature was then slowly lowered back to $T=0$ (dashed lines), but the configurations failed to order.}
    \label{fig:glassy}
\end{figure}

We describe some details of the Monte Carlo simulations of the melting transitions in the $J_4/U<0$ and $0>J_4/U>0.1$ regime.
As mentioned in the main text, we did not observe ordering for simulations starting from a disordered configuration at $J_4/U<0$, even as the temperature was decreased to $T=0$.
\autoref{fig:glassy}(a) shows an example of an ordered theoretical brick-wall monocrystal, where the blue lines mark the $J_4$ lines.
In comparison, \autoref{fig:glassy}(b) shows a typical example of a configuration from a simulation at $J_4/U=-0.1$ and $T/U=0$, which failed to order nematically and instead had many defects (purple lines) which prevented the system from ordering.
A similar situation arose for the $\sqrt{3} \times \sqrt{3}$ phase.
\autoref{fig:glassy}(c) shows a monocrystal where all trimers point in the same direction.
On the other hand, \autoref{fig:glassy}(d) shows a typical configuration from a simulation starting from a disordered configuration at $J_4/U=0.01$ and $T/U=0$ that failed to order.
Here, the different trimer colors mark different $\sqrt{3} \times \sqrt{3}$ domains.

To determine an upper bound to the brick-wall melting temperature despite the difficulty with simulation, we slowly heated up ordered configurations to find the (hysteretic) melting temperature.
\autoref{fig:glassy}(e) shows one such procedure performed at $J_4/U=-0.1$.
First, the simulation is initialized with a classical ground state (such as \autoref{fig:glassy}(a)).
The temperature is set to $T=0$, then slowly increases over the course of the simulation.
During the simulation, the structure factor indicating the brick-wall phase $S(q=M)$ is calculated, which decreases as the temperature increases.
We scaled the structure factor by $L^2$ to compare different system sizes.
At $T/U\approx 0.1$, the structure factor reaches its minimum value for all system sizes considered, indicating a completely disordered configuration.
We estimated the upper bound to the transition temperature to be when the structure factor reached half of its maximum value ($S(q=M)/L^2 \approx 0.05$), in this case, at approximately $T/U=0.085$.
After reaching $T/U=0.2$, the system was then cooled back to $T=0$.
However, the configurations were unable to reorder, even at $T/U=0$, due to the trimer model's strongly frustrated geometry.
This procedure was repeated for all points marked with pentagons in Figure 2(a) of the main text.

\widetext
\bibliographystyle{apsrev4-2}
\bibliography{refs}